\documentclass{JINST}
\usepackage{amsmath,amsfonts,amssymb,fontenc,times,mathptmx,graphicx}
\title{Improvements of Track Fitting with Well Tuned Probability Distributions for Silicon Strip Detectors}
\author{Gregorio Landi$^a$\thanks{Corresponding
author.}~,   Giovanni E. Landi$^b$\\
\\
\llap{$^a$} Dipartimento di Fisica e Astronomia,
Universita' di Firenze\\
Largo E. Fermi 2 50125 Firenze Italy\\
and INFN, Sezione di Firenze,
Firenze,Italy\\
E-mail: \email{landi@fi.infn.it}\\
\\
\llap{$^b$} UBICA s.r.l.,\\
Via S. Siro 6/1,\\
Genova, Italy.\\}

\abstract{
The construction of a well tuned probability distributions is illustrated in synthetic way,
these probability distributions are optimized to produce a faithful realizations
of the impact point distributions of
particles in silicon strip detector.  Their use
for track fitting shows a drastic improvements of a factor two, for the low noise
case, and a factor three, for the high noise case,
respect to the standard approach. The tracks are well reconstructed
even in presence of hits with large errors,
with a surprising effect of hit discarding.
The applications illustrated are simulations of the PAMELA tracker,
but other type of trackers can be handled similarly.
The probability distributions are calculated for the center of gravity algorithms,
and they are very different from gaussian probabilities.
These differences are crucial to accurately reconstruct
tracks with high error hits and to produce the effective discarding of the too noisy hits (outliers).
The similarity of our distributions with the Cauchy distribution forced us
to abandon the standard deviation for our comparisons and instead use
the full width at half maximum.
A set of mathematical approaches must be developed for these applications,
some of them are standard in wide sense, even if very complex.
One is essential and, in its absence, all the others are useless.
Therefore, in this paper, we report the details of this critical approach.
It extracts physical properties of the detectors, and allows the insertion of the
functional dependence from the impact point in the probability distributions.
Other papers will be dedicated to the remaining parts.}
\begin{document}

\section{Introduction }

Arrays of micro-strips are fundamental parts of almost
all the recent high energy physics experiments~\cite{hartmann}, their excellent position
resolutions are essential for track recognition.
Very sophisticated algorithms are developed to reconstruct the tracks, they extract
all the information released by particles crossing the sensitive area.
The complexity of these systems is astounding, special efforts are dedicated to the associations
of hits to tracks and track selection in an environment with
high noise and fake hits~\cite{CMS}~\cite{CMS1}.
The track reconstructions are always performed with $\chi^2$ minimization (often indicated as least squares)
or Kalman filters. Each of these methods assumes, in an implicit or explicit way,
identical Gaussian probability distribution functions (PDF) on large set of sensor arrays,
or  it is invoked, as a weaker justification, the optimality of the least squares among the
linear methods (Gauss-Markov theorem).
The advantages of these assumptions are evident, few details needed,
linear equations to solve and, in any case, acceptable solutions obtained.
The selection of  Gaussian models and the connected linear forms  is an obliged step
for such huge detectors.
Deviations from pure gaussian model are considered in ref.~\cite{fru,fru1}.
In those papers, the PDF are approximated as sums of gaussians, a type for the cores and
another for the tails of the distributions.
The method of gaussian sums can be tailored to conserve the main part of the Kalman filters
and their linearity. The resulting increase in resolution open the road to
explorations of more realistic forms with the possibility of further fit improvements.
This work introduces very specialized types of
PDF for minimum ionizing particle (MIP). These PDF are finely tuned
to contain many statistical properties of the MIP  on silicon micro-strip sensors.
Essential details of the hit-reconstruction algorithms and detector physics
are inserted in the mathematical expressions used in track fitting,
and our simulated fits show substantial improvements respect to the least squares method.
Hence, any move toward more realistic PDF has an evident gain.
Our PDF deviate strongly from gaussian PDF or gaussian sums, their
unusual forms are imposed by the
non-linearities of the most
used hit-reconstruction algorithms: the center of
gravity (COG) defined as $(\sum_i E_i l_i)/\sum_i E_i$ (sometimes called "centroid").
The weighted average (another name for the COG) of the strip
positions ($l_i$), with weights depending from the strip signals
($E_i$), gives a hit resolution much better than the strip size.
The spreading of particle signal in
few nearby strips is a key element of this improvement, and in some devices
the spreading is enhanced with appropriate
cross-talk. The importance of the hit
reconstruction is evident: better positions give
better fits.

The gain, produced by the signal spreading, can be maximized
with different COG algorithms with a different number of strips.
These algorithms can be tuned at any specific situation and are able to keep the noise to a minimum
maintaining the resolution.
Each COG algorithm has its own set of systematic errors and
very different PDF, it is important to operate with
similar algorithms in similar set of events.

Our detector model is
the PAMELA tracker~\cite{PAMELA} for MIP events.
The PAMELA sensors are double-sided silicon strip detectors~\cite{aleph}
of the type we used in the L3 micro-vertex detector~\cite{L3}.
Each side has strips with different properties. On one side, the read
out electronics is applied to all the strips, this side has properties
similar to conventional micro-strip detectors.  On the other
side, one strip each two is connected to the readout.
The unconnected strip is left to a floating potential
and it spreads the charge, released around it, to nearby readout strips.
This side will be called floating strip side.
Thus, we have to study two
sensor types  with very different signal to noise
ratio and charge distributions.

Given the complex development needed to complete the work, it will be split in various part and
published separately. Here all the details that conduct to our PDF will be neglected, in a wide
sense these developments are standard~\cite{prob} even if very complex.
Section two describes few properties of our PDF and its non gaussian form.
In section three, a mathematical tool will be developed, it defines
an average energy collected by a strip at any impact point
This is a key element that allows the association of a most
probable set of impact points to each hit.
Section four and five are dedicated
to the simulation of tracks on this two-sided silicon detector,
the floating strip side with low noise and
large charge spreading, and the normal strip side with
high noise and small charge spreading.

\section{Probability distributions}

The PDF for the COG algorithms are fixed by the
details of the COG calculation; the
number of strips involved and the rules of the strip
selection are the most important. A detailed discussion of these aspects is reported in
in ref.~\cite{landi01,landi02,landi03}, a special attention is devoted to
the systematic errors introduced by the discretization of the charge distributions.
In ref.~\cite{landi01,landi02} we handled the signals with the formalism of large use
for the Shannon sampling theorem (or with the present naming convention: the
Whittaker Kotel'nikov Shannon sampling theorem~\cite{jerry}). In fact,
a modern particle detector performs a sampling of the incident signals,
and the natural methods to treat them are just those developed to reconstruct
analogical signals from their sampled forms. This well grounded
formalism  allows us to demonstrate properties that will be encountered
in the following.
One property, discussed in ref.~\cite{landi01,landi03}, is the effect of strip
suppression. The limitation on the number of strips
is very beneficial for the noise reduction, but these suppressions
add typical systematic errors. The COG algorithms with an even number of
strips have a set of forbidden values corresponding to the strip center,
those with odd numbers have forbidden values around the strip borders.
For $n$-strip COG, the amplitudes of the forbidden regions are proportional to the
signal amplitudes escaping the $n-1$ strips. Realistic PDF must accurately reproduce them.

\subsection{Probability distribution for two strip COG}

In ref.~\cite{landi01,landi03} no attention was given to the signal
fluctuations, now this aspect is central: the signals are (gaussian) random
variables and (non-gaussian)  random variables are the results of the COG algorithms
operating on them.
The two strip COG will be our starting point, this algorithm
has the lowest noise and is well suited around the orthogonal
incidence (our selected direction in the simulations).
As we anticipated, the extraction of an usable
form of the PDF for the two strip COG is a very complex development,
and we skip now all these huge details, we recall simply some key points that are
essential in the following. Identically we skip the heavy details of the
PDF for three, four and five strip COG, even if the results for the
three strip COG will be recalled.

By definition, the weights of a
COG algorithm are ratio of different combinations of independent random variables,
hence the PDF of the COG values share some similarity with the PDF given by the ratio of two
independent random variables. For example, these PDF go to infinite as a
Cauchy distributions when gaussian random variables are implied. Therefore,
given our assumption (as usual) of the gaussian PDF for the strip signals,
similar tails must be expected even in our case.
The two strip COG algorithm is more elaborate than
the ratio of two random variables, and, even if the gaussian integrals in the PDF
have not a closed form, the characteristic heavy tails
of the Cauchy distributions are isolated from the very beginning.
Similar results are encountered for the PDF of COG
with an higher number of strips, no closed form for the gaussian integrals,
and Cauchy-like tails.

For our heavy use, it is impossible to handle
numerical integrations, and  we are forced to find approximate
analytical forms able to give excellent reproductions
of the PDF. The most complete and accurate of these forms are very long,
almost a printed page for each of them, thus we are obliged to publish
them separately. We will synthesize the properties of the PDF
$P_{x_{g2}}(x) $ for the two strip COG (indicated by $x_{g2}$) with:
\begin{equation}\label{eq:equation_0}
    P_{x_{g2}}(x)=\frac{F(a_1,a_2,a_3,\sigma_1,\sigma_2,\sigma_3,x)}{x^2}\,.
\end{equation}
The function $P_{x_{g2}}(x)$, for the COG value $x$,  depends from six constants, the three
$a_1,a_2,a_3$ are the mean values of the gaussian PDF. They are the noiseless
charges collected by the strips. The other three constants $\sigma_1,\sigma_2,\sigma_3$
are the amplitudes of the gaussian noise.
The parameters of central strip have index $2$, and the indices $1$ and $3$ indicate respectively
the parameters of the right and left strip.
For $x$ going to infinite, the $x$-dependence in the function $F$ goes to one,
hence $P_{x_{g2}}(x)$ is a slow decreasing function
similar to a Cauchy distribution and sharing with this the infinite variance.
The divergence of the variance produces long tails in a finite
set of data, these tails rule out the use of the root mean
square error as an useful parameter to compare the error distributions.
For this we will use the full width
at half maximum (FWHM) that is well defined even for Cauchy-like distributions.

\begin{figure}[h]
\begin{center}
\includegraphics[scale=0.8]{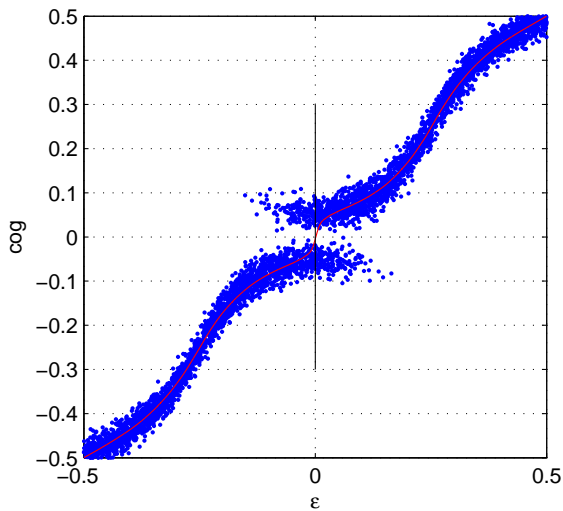}
\caption{\em Scatter plot of $x_{g2}$ in function of its impact
point $\varepsilon$ for a two strip COG algorithm.
The continuous red line is $x_{g2}(\eta_2)$ extracted from the simulated data. The back line indicates the $\varepsilon$-value
used to illustrate the form of $P_{x_{g2}}(x)$
}\label{fig:figure_1}
\end{center}
\end{figure}

The analytical approximations
we derived for $P_{x_{g2}}(x)$ (and all the other
with 3, 4, 5 strips) have very small differences respect
to those given by numerical integrations, around $10^{-5}$ or less.
One of these expressions is illustrated by its comparison with
hit simulations of floating strip detector~\cite{landi03}.
The incidence angle is orthogonal to the detector plane,
and the MIP charge spreads on few nearby strips.
The non gaussian aspect of the
COG probability is more evident for the impact point  $\varepsilon\approx0$
indicated with a vertical black line in fig.~\ref{fig:figure_1}.
At this orthogonal incidence, the set of forbidden
COG values around zero  is  reduced  to a
minimum. Even if the algorithm accumulates
its largest position error in this region, it globally
remains much better than the three strip COG that is free of this gap.
The addition of one strip drastically increases the
noise and degrades the position reconstructions,
for this reason the two strip COG is preferred.

The continuous red line of fig.~\ref{fig:figure_1} is
the result of the  position algorithm
called $\eta_2$ in ref.~\cite{landi03} and it is a generalization
of the $\eta$-algorithm of ref.~\cite{belau}. The $\eta_2$-algorithm is derived
from the COG distributions, it
suppresses appreciably the COG systematic error discussed in ref.~\cite{landi01} and
places the red line in the maximum of the $\{\varepsilon,x_{g2}\}$ density distribution.
On the contrary, a pure COG algorithm assumes its value
as an estimation of $\varepsilon$, thus its relation with $\varepsilon$ is a straight-line
from $\{-0.5,-0.5\}$ to $\{0.5,0.5\}$,  appreciably different from the point distribution and the red line.
The COG straight-line is not reported in fig.~\ref{fig:figure_1} because we will never use the simple COG for
position reconstructions.

For the $x_{g2}$ distribution, fig.~\ref{fig:figure_1} shows two branches
around $\varepsilon\approx0$, these branches are  produced by
the noise that increases the signal of the left strip for
$\varepsilon>0$ or increases the signal of the right strip for $\varepsilon<0$.

\begin{figure}[h]
\begin{center}
\includegraphics[scale=0.7]{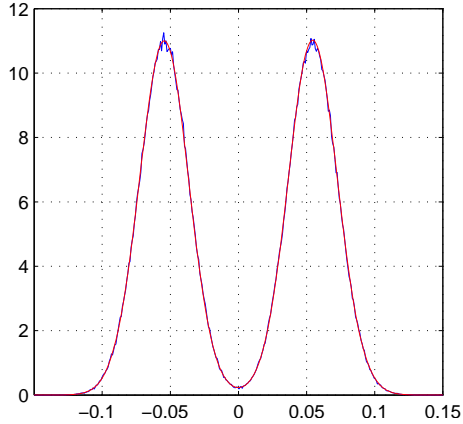}
\caption{\em Probability distribution (blue line) for $x_{g2}$ at fixed $\varepsilon=0.0001$ from a set of
simulated data  ( $10^6$ events of fixed $a_1,a_2,a_3$) fully overlapped by our PDF (red line). }
\label{fig:figure_2}
\end{center}
\end{figure}

A sample of $P_{x_{g2}}(x)$ is plotted in fig.~\ref{fig:figure_2}, the set of
data is generated with the constants  $a_1,a_2,a_3$, the noiseless strip
signals of the selected $\varepsilon$-value, and identical standard deviations
$\sigma_1$, $\sigma_2$ and $\sigma_3$ of their gaussian noise
(4 ADC counts as reported in ref.~\cite{vannu}).
The line of $P_{x_{g2}}(x)$ overlaps completely the  normalized
$x_{g2}$-histogram of the data sample, confirming
the quality of these analytical approximations and of our PDF.
This plot reproduces the PDF just in the critical region of the
two strip COG where a forbidden region of $x_{g2}$-values is present
in the noiseless algorithm. The noise dresses the two noiseless spikes
giving them the form of two shifted gaussian-like curves, but
the tails are Cauchy-like as in eq.~\ref{eq:equation_0}.

The most difficult testing point for our analytical $P_{x_{g2}}(x)$
is around $\varepsilon=0$ for MIP direction
inclined of $20^\circ$ respect to the orthogonal incidence,
here the forbidden gap is very large ($\approx 0.8\tau$)
with two narrow bumps, but even in
this case the $P_{x_{g2}}(x)$  gives a perfect
match with the data histogram.
At these angles, the three strip COG is correct selection,
but our analytical form must work even here.

The simulated data are produced with uniform ionization
along the particle path, the charge diffuses in a
standard way up to the collection by the
readout electrodes.
This model neglects the non-uniform ionization
along the particle path, but its insertion does not
modifies fig.~\ref{fig:figure_2}. The fluctuation
of the charge release looks well contained in the Landau
distribution of the total collected charge.
\section{The introduction of the impact point}

The PDF of eq.~\ref{eq:equation_0} are defined for a fixed $\varepsilon$, as shown in
fig.~\ref{fig:figure_2}, and give the corresponding $x_{g2}$ distributions.
This is the maximum that the probability theory can give, but, in spite of the work invested,
these forms are  useless for track reconstruction, any further
improvement must go in depth into the sensor physics.
In fact, useful PDF must associate sets of $\varepsilon$-values to
the measured COG for each hit, the COG is the information given by the detector.
The standard gaussian model attributes identical set of $\varepsilon$-distributions
at each hit. This assumption looks very improbable just
observing fig.~\ref{fig:figure_1} for small $x$-values.
If we cut fig.~\ref{fig:figure_1} with constant
COG lines, the thickness of the blue point distribution is very different for each selected value.
The thicknesses  become negligible around zero and increase drastically
above and below zero, this region has the
strongest deviations from constant gaussian PDF.

To complete our analytic approach to a well
tuned PDF, we need the probability of $\varepsilon$
at fixed $x$ along lines orthogonal  to the black
line of fig.~\ref{fig:figure_1}. Therefore  eq.~\ref{eq:equation_0}
must be extended to contain the $\varepsilon$ functional dependence.
No $\varepsilon$-dependence can be supposed for the
strip noises $\sigma_1,\sigma_2,\sigma_3$,
they depend from the quality of the strip (noisy strips,
average strips, etc.), and are obtained from the pedestal runs.
On the contrary, the energies $a_1,a_2,a_3$ surely
depend from the impact point $\varepsilon$, and a
method must be defined to extract this functional dependence from the data.

Along the particle path, the ionization fluctuates
projecting the fluctuations on the collecting strips,
hence, well defined functions
$a_1(\varepsilon),a_2(\varepsilon),a_3(\varepsilon)$ do not exist.
They must be defined as averages over the signal fluctuations.
A procedure could be the use of the average signal
distribution defined in ref.~\cite{landi01,landi03} with
the $\mathrm{d} x_{gn}(\varepsilon)/\mathrm{d}\varepsilon$.
But, the functions $a_1(\varepsilon),a_2(\varepsilon),a_3(\varepsilon)$
require the convolution of the signal distribution with an interval
function large as its period (a strip), and this convolution
is always a constant function. A different path, to avoid this trivial result, must be found.

\subsection{Brief synthesis of the  $\eta$-algorithm}

In the following, we will amply use our generalizations of the
$\eta$-algorithm especially in the extraction
of the $\{a_k(\varepsilon)\}$-functions, hence
a rapid synthesis of these extensions is in order.
A key point for the effectiveness of the original $\eta$-algorithm
is the compensation of the
non-uniformity of the two strip COG histograms.
For the sake of precision, the original algorithm dealt with
a special combination of two strip signals called $\eta$-function,
but the $\eta$-function can be transformed in a two strip COG,
this justifies our constant reference to the COG.

It is easy to observe that a large set of MIP , crossing almost
uniformly a detector,  produces non uniform COG histograms,
this means that the COG algorithm has privileged outputs
(systematic errors). A good position reconstruction
algorithm must produce uniform histograms, and this is just
the result of the $\eta$-algorithm.
The derivation of ref.~\cite{belau} limited
the algorithm to two strips and to symmetrical signal
distribution, this very improbable condition renders
problematic its use for generic geometry.
In ref.~\cite{landi03} we demonstrated how to extend the
$\eta$-algorithm for any asymmetric signal distribution
and any number of strips, for this we introduce
the notation $\eta_j$-algorithms ($j$ is the number of strips implied).
In its simplest form, our extended $\eta_j$-algorithms are
defined as :
\begin{equation*}
    \eta_j(x_{gj})=x_{gj}+C_0+\int_{-\tau/2}^{x_{gj}}\big[\tau \Gamma_j^p(x_{gj}')-1\big]\,\mathrm{d}x_{gj}'\,.
\end{equation*}
Where $\tau$ is the strip length, $\Gamma_j^p(x_{gj}')$ is the periodic PDF
for  the $j$-strip COG $x_{gj}'$, it is  given by
a uniform illumination over the detector plane
and normalized to one on a strip (essentially a normalized histogram divided by
the bin size and shifted by a set of strip length).
Therefore, the integral term is periodic and can be expressed as a Fourier Series.
The constant $C_0$ is $0$ for symmetric
signal distributions. The general  expression for $C_0$ is reported in
ref.~\cite{landi03,landi05} with other details about $\eta_j(x_{gj})$
(indicated there as $\varepsilon_j(x_{gj})$ and $\Gamma^p(x_{gj})$ ).
These extensions of the $\eta_j$-algorithms were successfully verified in a dedicated test beam
experiment~\cite{vannu} with the use of a special set-up. The test beam evidenced
even another subtle systematic error whose correction was
discussed in ref.~\cite{landi05}.
All these corrections are crucial for our  PDF,
the final result must produce the maximum of the probability just along the
continuous red line of fig.~\ref{fig:figure_1} where it is
evident the largest COG population. Any systematic error of
the $\eta_2$-algorithm will introduce
a shift from this optimal position and a corresponding error in the
fitted tracks with any type of fitting algorithm. The consistency of all our procedures
will be verified in the following.

\subsection{The definition of the strip energies $a_1(\varepsilon),a_2(\varepsilon),a_3(\varepsilon)$}

For the extraction of the functions $a_1(\varepsilon),a_2(\varepsilon),a_3(\varepsilon)$,
the appropriate formalism is that of ref.~\cite{landi01} with extensive use of
the sampling theorems~\cite{jerry}. But, due to the central role of these functions as our
key to go in depth in the sensor physics, we
will follow a simpler approach where the sole properties of the Fourier
transform and the Fourier Series are implied.

Let us define our notations.
The signal collected by a strip is obtained from the convolution of the strip
response function $g(x)$ with the average charge
distribution $\varphi(x-\varepsilon)$. The function
$\varphi(x-\varepsilon)$ is defined to have its COG coinciding
with the impact point $\varepsilon$ of the MIP, and
normalized to one.  This position fixes $\varepsilon$
to be the COG of the average primary ionization segment~\cite{landi03}.
The average strip signals are given by the
convolution:
\begin{equation}\label{eq:equation_1}
    f(x-\varepsilon)=\int_{-\infty}^{+\infty}g(x-x')\varphi(x'-\varepsilon)\,\mathrm{d} x'\,.
\end{equation}
For any $n\in \mathbb{Z}$ and $\tau$ the strip length, the value of the function
$f(x-\varepsilon)$ at $x=n\tau$ gives the signal collected by the strip
with its center in $n\tau$ (sampling at $n\tau$).
The origin of the reference system is the center of strip with
the maximum energy (~with the index 2~), the other two strip centers are to the right and to the left with
position $+\tau$ and $-\tau$ (~indices 1 and 3~). The short range
of $\varphi(x)$ and $g(x)$ gives very few non zero $f(n\tau-\varepsilon)$, hence all our
infinite sums will be always convergent. In the following, the strip length $\tau$ is always taken equal to one,
its indication is reported to assure the right dimensions of the equations.

The functions
$a_1(\varepsilon),a_2(\varepsilon),a_3(\varepsilon)$ are expressed as:
\begin{equation}\label{eq:equation_2}
\begin{aligned}
    &a_1(\varepsilon)=f(\tau-\varepsilon)\\
    &a_2(\varepsilon)=f(-\varepsilon)\\
    &a_3(\varepsilon)=f(-\tau-\varepsilon)\\
    &|\varepsilon|\leq \frac{\tau}{2}
\end{aligned}
\end{equation}
With the shifts of eq.~\ref{eq:equation_2}, the function $f(-\varepsilon)$
produces all the functions $a_J(\varepsilon)$.

We will prove that: {\em if a generic signal distribution is a convolution with
an interval function, with the size of the strip or any of its multiples, the sum of the signal collected
by all the strips is independent
from the impact point for any type of strip-loss}.

Let us consider all the strips at intervals $T=N\tau$ (~$N$-multiple of  $\tau$~).
These strips are the sole utilized, all the others are neglected. This
selection generates an effective drastic loss that allows the
exploration of the tails of the function $f(x)$.
The integration of the function $f(kT-\varepsilon)$ (~for $k\in \mathbb{Z}$~)
with an interval function $\Pi(x/T)$ (~$\Pi(x)=1$ for $|x|\leq 1/2$ and $\Pi(x)=0$ for $|x|>1/2$~) of size
$T$ will be our starting point to reconstruct $f(x)$:
\begin{equation}\label{eq:equation_3}
    h(kT-\varepsilon_f)=\int_{-\infty}^{+\infty}\Pi(\frac{\varepsilon_f-\varepsilon}{T})f(kT-\varepsilon)\,\mathrm{d}\varepsilon\,.
\end{equation}
For future manipulations is better to rearrange eq.~\ref{eq:equation_3} as a convolution:
\begin{equation}\label{eq:equation_3a}
    h(kT-\varepsilon_f)=\int_{-\infty}^{+\infty}\Pi(\frac{kT-\varepsilon_f-\zeta}{T})f(\zeta)\,\mathrm{d}\zeta\,.
\end{equation}
The function $\sum_{k\in \mathbb{Z}} h(kT-\varepsilon_f)$ is obtained
by shifting copies of the function $h(-x)$
for all the intervals $kT$, this produces a periodic function
in $\varepsilon_f$ with period $T$. As we said, the short range of $h(-x)$ assures the convergence
of the infinite sum. Let us show that it is constant for
any $\varepsilon_f$ as stated previously.
Due to periodicity, $\sum_{k\in \mathbb{Z}} h(kT-\varepsilon_f)$ can be expressed with a Fourier Series and
the Poisson identity~\cite{libroFT2,libroFT} gives this form of Fourier Series:
\begin{equation}\label{eq:equation_4}
    \sum_{k=-\infty}^{+\infty}\,h(kT-\varepsilon_f)=\frac{1}{T}
    \sum_{L=-\infty}^{+\infty}\,\mathrm{e}^{\mathrm{i}  2\pi\,L\,\varepsilon_f/T} H(-\frac{2\pi\,L}{T})\,.
\end{equation}
Where $H(-2\pi\,L/T)$ is the Fourier Transform of $h(x)$ in the points $-2\pi\,L/T$.
For the convolution theorem, the
Fourier Transform of $h(x)$ is the product of the Fourier transforms of
$f(x)$ and $\Pi(x/T)$ respectively $F(\omega)$ and $2\sin(\omega T/2)/\omega$.
Due to eq.~\ref{eq:equation_1}, $F(\omega)$ is the product
of $G(\omega)\Phi(\omega)$ the Fourier transform of $g(x)$ and $\varphi(x)$.
The term $2\sin(\omega T/2)/\omega$  is zero for
all  $\omega=2\pi\,L/T$ with $L\neq0$ and equal to $T$ for $L=0$.
Therefore, all the terms depending from $\varepsilon_f$
are zero and $\sum_{k\in \mathbb{Z}} h(kT-\varepsilon_f)$  is constant as we stated.
Due to our definition of $\Phi(0)=1$, eq.~\ref{eq:equation_4} becomes:
\begin{equation}\label{eq:equation_5}
    \sum_{k=-\infty}^{+\infty}\,h(kT-\varepsilon_f)=G(0)=\int_{-\infty}^{+\infty}g(x)\,\mathrm{d} x\,.
\end{equation}
This result can be obtained even with a different path. With another change of variable in eq.~\ref{eq:equation_3a},
the sum $\sum_{k\in \mathbb{Z}} h(kT-\varepsilon_f)$ reduces to an
integral from $-\infty$ to $+\infty$ of eq.~\ref{eq:equation_1} and the convolution theorem gives eq.~\ref{eq:equation_5}.

The COG with the $h(kT-\varepsilon_f)$ is:
\begin{equation}\label{eq:equation_6}
    x_g(\varepsilon_f)=\frac{\sum_{k=-\infty}^{+\infty}\,h(kT-\varepsilon_f)kT}{\sum_{n=-\infty}^{+\infty}\,h(nT-\varepsilon_f)}\,.
\end{equation}
Isolating the periodic part and inserting eq.~\ref{eq:equation_5}, it becomes:
\begin{equation}\label{eq:equation_7}
    x_g(\varepsilon_f)-\varepsilon_f=\frac{1}{G(0)}\sum_{k=-\infty}^{+\infty}\,h(kT-\varepsilon_f)(kT-\varepsilon_f)
\end{equation}
that can be converted in:
\begin{equation}\label{eq:equation_8}
   \lim_{\omega\rightarrow 0} i\frac{\mathrm{d}}{\mathrm{d}\omega}\sum_{k=-\infty}^{+\infty}\,\mathrm{e}^{-\mathrm{i}\omega(kT-\varepsilon_f)}h(kT-\varepsilon_f)=
   G(0)(x_g(\varepsilon_f)-\varepsilon_f)\,.
\end{equation}
Again the Poisson identity gives a generalized form of eq.~\ref{eq:equation_4}:
\begin{equation}\label{eq:equation_9}
    \sum_{k=-\infty}^{+\infty}\,\mathrm{e}^{-\mathrm{i}\omega(kT-\varepsilon_f)}h(kT-\varepsilon_f)=
    \frac{1}{T}\sum_{L=-\infty}^{+\infty}\,H(\omega-\frac{2\pi\,L}{T})\mathrm{e}^{\mathrm{i}\,2\pi\varepsilon_f\,L/T}
\end{equation}
As defined in eq.~\ref{eq:equation_4}, the function $H(\omega-2\pi\,L/T)$ is the product of all the
convolved functions, and its derivative is a sum of terms with a derivative of single factor.
For $\omega\rightarrow 0$ and $L\neq 0$, all the terms without the
derivative of $2\sin(\omega T/2)/\omega$ (the Fourier transform of $\Pi(x/T)$) are zero.
For $L=0$ and $\omega\rightarrow 0$, the convolution theorem for the first momenta defines
$\mathrm{d} H(\omega)/\mathrm{d} \omega$ to be the sum of terms proportional to the first momenta of the
convolved functions. The interval function $\Pi(x/T)$ is symmetric and its first momentum is zero.
The first momentum of $\varphi(x)$ is zero by our definition. If $g(x)$ is symmetric even
its first momentum is zero, otherwise it will give the only non zero term.
For asymmetric $g(x)$, the shift $\delta_g$ of the strip COG respect to the strip center must be added.
In this general case  $x_g(\varepsilon_f)-\varepsilon_f$ becomes:
\begin{equation*}
    (x_g(\varepsilon_f)-\varepsilon_f-\delta_g)G(0)=
    -\mathrm{i}\sum_{L=-\infty,L\neq 0}^{+\infty}\,F(-\frac{2\pi\,L}{T})\frac{T(-1)^L}{2\pi\,L}\mathrm{e}^{\mathrm{i}\,2\pi\varepsilon_f\,L/T}
\end{equation*}
Substituting $(-1)^L=\mathrm{e}^{\mathrm{i}\pi\,L}$, the derivative of $x_g(\varepsilon_f)$ respect to $\varepsilon_f$ gives:
\begin{equation}\label{eq:equation_10}
    \frac{\mathrm{d}\,x_g(\varepsilon_f)}{\mathrm{d}\varepsilon_f}=1+
    \frac{1}{G(0)}\sum_{L=-\infty,L\neq 0}^{+\infty}\,F(-\frac{2\pi\,L}{T})\mathrm{e}^{\mathrm{i}\,2\pi(\varepsilon_f+T/2)\,L/T}\,.
\end{equation}
With the addition of the missing term $F(0)/G(0)$, the sum, in right hand side of eq.~\ref{eq:equation_10},
becomes a Fourier Series . But $F(0)/G(0)$ must be 1. In fact,
$F(0)$ is $G(0)\Phi(0)$, with $\Phi(0)=1$, thus  the first term of this side is just $F(0)/G(0)$:
\begin{equation*}
    \frac{\mathrm{d}\,x_g(\varepsilon_f)}{\mathrm{d}\varepsilon_f}=
    \frac{1}{G(0)}\sum_{L=-\infty}^{+\infty}\,F(-\frac{2\pi\,L}{T})\mathrm{e}^{\mathrm{i}\,2\pi(\varepsilon_f+T/2)\,L/T}\,.
\end{equation*}
Applying back the Poisson identity (defined in eq.~\ref{eq:equation_4} with our notations), it can be written as:
\begin{equation}\label{eq:equation_11}
    \frac{\mathrm{d}\,x_g(\varepsilon_f)}{\mathrm{d}\varepsilon_f}=\frac{T}{G(0)}\sum_{n=-\infty}^{+\infty}\,f(nT-\varepsilon_f-\frac{T}{2})\,.
\end{equation}
The factor $T$ is absent in eq.~\ref{eq:equation_10} and must be inserted.
Equation~\ref{eq:equation_11} underlines the possible presence of tails interference
(called "aliasing" in signal theory) if the range of $f(x)$
is larger than $T$. The aliasing can be limited selecting a reasonable
period $T$ to reproduce correctly these tails. In the absence of aliasing,
a section of amplitude $T$ and $n=0$ of eq.~\ref{eq:equation_11} is given by:
\begin{equation}\label{eq:equation_12}
    \frac{\mathrm{d}\,x_g(\varepsilon_f)}{\mathrm{d}\varepsilon_f}=\frac{T}{G(0)}f(-\varepsilon_f-\frac{T}{2})\,,
\end{equation}
that easily produces the $a_j(\varepsilon)$ of eq.~\ref{eq:equation_2}
with the appropriate shifts.

Due to eq.~\ref{eq:equation_10},  the normalization of
$\mathrm{d} x_g(\varepsilon_f)/\mathrm{d}\varepsilon_f$ on a period $T$ is just $T$,
thus the remaining factor is normalized to one in any case.
The normalization of $g(x)$ is $\tau$ (one for our conventions) in
absence of loss, and decreases with the loss. The fixed normalization of
$\mathrm{d} x_g(\varepsilon_f)/\mathrm{d}\varepsilon_f$
does not allow the extraction of the average strip efficiency $G(0)$ in presence of a loss.

\subsection{The sliding window and its simulation}

The form of eq.~\ref{eq:equation_3} suggests an experimental procedure to extract
the functions $\{a_j(\varepsilon)\}$. A rectangular window, as wide
as the period T and with a side parallel to the strips, is required. The window slides on the detector
and, at each step, collects a fixed amount of signal from the events contained in its interior.
With very small steps
and collecting a large number of events, this set of data can approximate a convolution
with a function $\Pi(x)$. The use of eq.~\ref{eq:equation_12} on the COG calculated
with eq.~\ref{eq:equation_6} gives the unknowns.

In the absence of this special setup, the data collected in a
standard test beam can be a valid substitution, but the result could contain some artifacts.
The following approximation of equation~\ref{eq:equation_3} can
substitute the sliding window:
\begin{equation}\label{eq:equation_12a}
    \int_{-\infty}^{+\infty}
\Pi(\frac{\varepsilon_f-\varepsilon}{T})f(kT-\varepsilon)\,\mathrm{d}\varepsilon\approx
\sum_{\varepsilon_i=\varepsilon_f-T/2}^{\varepsilon_f+T/2}f(kT-\varepsilon_i)\Delta_i
\end{equation}
The terms $f(kT-\varepsilon_i)$ are the charges collected
by the selected strips at distance $kT$, their values fluctuate
due to the mechanism of charge release. In the simulation,
careful averages of eq.~\ref{eq:equation_12a} and the use of
the impact points are able to produce $a_i(\varepsilon)$
identical to the starting ones. In a test beam data or in a running experiment, the
impact points are not available and the use of estimated
values is obliged. The COG values must be excluded due to
a linear correlation with the strip signals, in this case the
$\{a_i(\varepsilon)\}$-functions turn out to be almost identical triangular
functions. The outputs of the $\eta_2(x_{g2})$ and $\eta_{3}(x_{g3})$
algorithms give good results in the simulations. Their error
distributions introduce small
artifacts and distortions in the function $f(x)$,
but almost all can be corrected with an accurate
research of their origins. The final result shows
very smooth shapes unexpected for a Monte Carlo
integration.
After acquiring confidence with
the method in the simulations, we utilize it with
test beam data of ref.~\cite{vannu}.

The test beam data are spread over a large number of different strips, but it is
easy to collect the hits to have an
identical maximum signal strip, this collection
simulates a uniform illumination on that strip.
We need a uniform distribution over a larger
strip set, fifteen strips are used in our reconstruction.
An integer random number  is selected for each hit
and all the elements of the hit are shifted by a number
of strips corresponding to this integer. These translations
simulate a uniform hit population on the selected strips.
The sliding window (five strips wide in this test)
operates on this distribution, isolating the data
whose estimated ($\eta_2,\eta_3$)-positions are
contained in the window. This shuffling is done
many times and the results of the sliding window
operations are averaged.
The quality of the resulting functions $a_i(\varepsilon)$
can be tested by a confront with the histograms of $x_{g2}$
and $x_{g3}$ as illustrated in the following.

\subsection{Extended probability distributions}
The functions $a_1(\varepsilon),a_2(\varepsilon),a_3(\varepsilon)$ introduce
in our (a page long) PDF the impact point $\varepsilon$:
\begin{equation}\label{eq:equation_13}
    P_{x_{g2}}(x,E_t,\varepsilon)=\frac{F(a_1(\varepsilon),a_2(\varepsilon),a_3(\varepsilon),E_t,\sigma_1,\sigma_2,\sigma_3,x)}{x^2}\,.
\end{equation}
The $a_i(\varepsilon)$ are the fractions of the total
signal collected by each strip, and the total (noiseless) signal $E_t$ must be introduced
explicitly in eq.~\ref{eq:equation_13}. For simplicity, all the $\sigma_j$ are taken identical and dropped
from the notation, but it is evident the possibility to consider noisy strips.
In the application, each hit has the normalizing constant $E_t$ given by
the sum of the signals collected by the central strip and the two lateral ones.
This total signal contains noise, but nothing better is available. Now the function
$P_{x_{g2}}(x,E_t,\varepsilon)$, for each $E_t$, is a surface in the $\{x,\varepsilon\}$-plane.
The measured COG of the hit, the $x$-value, cuts on this
surface the $\varepsilon$-distribution that
will be used in the maximum likelihood search.

The normalization of $P_{x_{g2}}(x,E_t,\varepsilon)$ on $x$ for any $a_j$ and $E_t$
imposes to the marginal probability
$P_{x_{g2}}(\varepsilon)$ to be always equal
to one. This produces an effective uniform "illumination" that is consistent with
our assumptions.
The other marginal probability $H_{x_{g2}}(E_t,x)$ is connected to the histogram of $x_{g2}$.
In fact, the probability of  $x_{g2}=x$ for a
total event signal $E_t$ is given by the integration of
$P_{x_{g2}}(x,E_t,\varepsilon)$ on $\varepsilon$ over a reasonable range of $\varepsilon$-values.
The range of integration is not critical, it must cover all the $\varepsilon$-values that give
an appreciable contribution to the integral, and remain in the region where our $a_j(\varepsilon)$ are well defined.
We standardize this range to a two strip length.   The resulting $H_{x_{g2}}(E_t,x)$ must be calculated
for $x$ in the interval $-0.5<x\leq 0.5$, and averaged over the probability of $E_t$. This can be compared with the histogram of the data
to test the quality of the functions $\{a_j(\varepsilon)\}$. The histograms are very sensible to these
functions, and further improvement could be introduced.
As we said above, a lot of work has been dedicated to find analytical
approximations for the COG PDF. These approximate analytical expressions, even if very long,
are essential to speed up this comparison. Otherwise, each point of our calculated histogram would be obtained
by multidimensional numerical integrations.

To evaluate the results of our PDF, simulations are produced as near to the data as possible.
For this, the $\{a_j(\varepsilon)\}$ are extracted from the data of a test beam~\cite{vannu} with sensors
identical to those of the PAMELA tracker.
The detector was formed by five layers of two-sided sensors without magnetic field.
The slight differences among sensors are neglected and all the data of similar
sides are used together to increase the data sample.

\section{Floating strip side}

\begin{figure}[t]
\begin{center}
\includegraphics[scale=0.9]{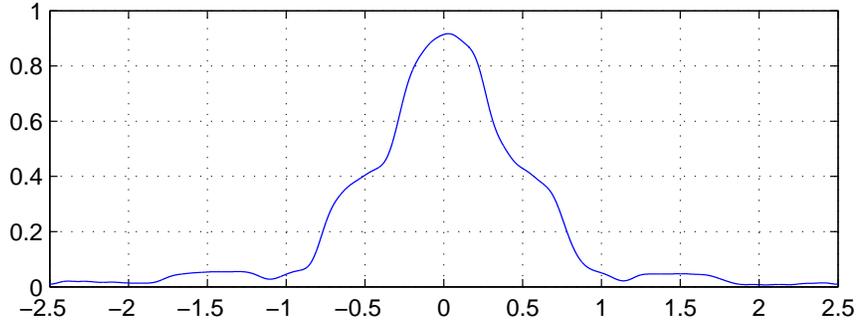}
\caption{\em  The reconstructed function $f(-\varepsilon)$ for the floating strip size
 }\label{fig:figure_3}
\end{center}
\end{figure}

This side has the lowest noise  (4 ADC counts) and the maximum of the charge released around 142 ADC counts.
The functional relations of $a_j(\varepsilon)$ and $\sigma_j$ in eq.~\ref{eq:equation_13} are scale invariant
and we are free to measure all them in
ADC counts with $x$ as a pure number in unity of strip length. The function $f(-\varepsilon)$ for this side is illustrated in
fig.~\ref{fig:figure_3}. Its aspect is similar to that obtained in ref.~\cite{infrared_laser} with a laser test,
surprising similar are the lateral crosstalk around $\pm 1.5$.
The dips around $\pm 1$ are small reconstruction artifacts. The shoulders around $\pm 0.5$ are typical of
the floating strips~\cite{landi03}, they spread the charges to nearby strips producing
the two lateral peaks of COG histograms.

\begin{figure}[b]
\begin{center}
\includegraphics[scale=0.55]{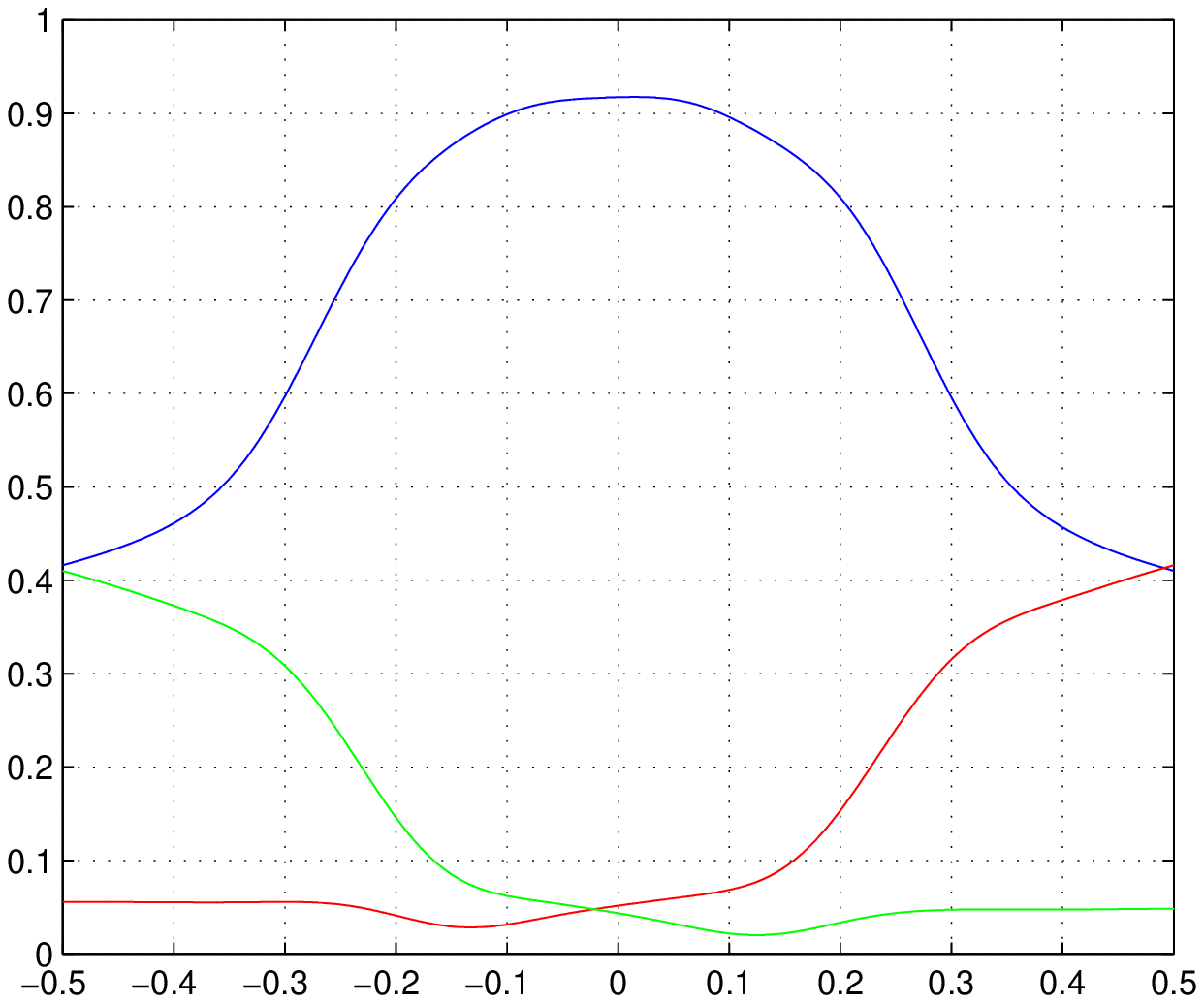}
\includegraphics[scale=0.51]{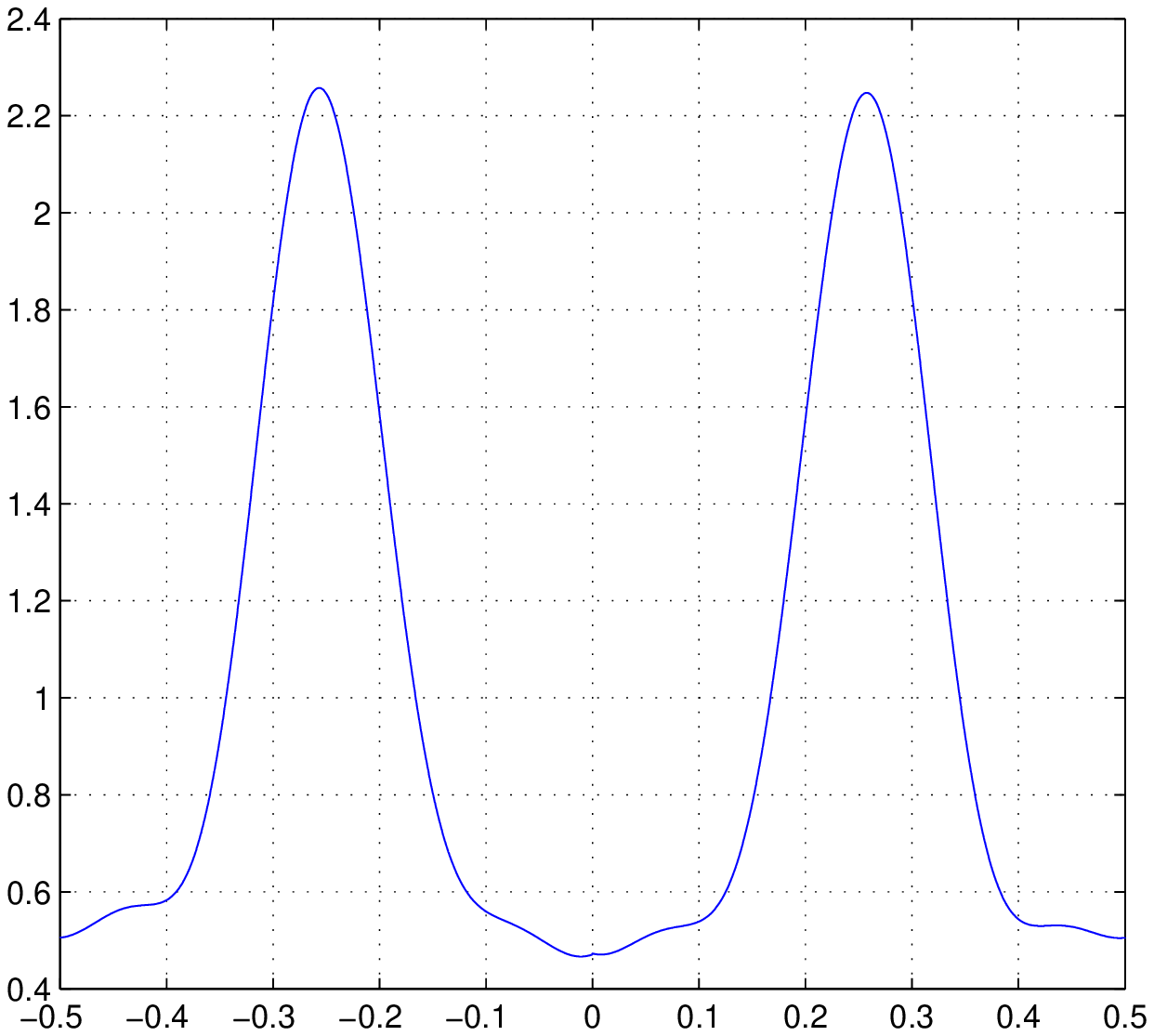}
\caption{\em To the right: the three energy fractions $a1(\varepsilon)$ (red),
$a2(\varepsilon)$ (blue), $a3(\varepsilon)$ (green). To the left: the average form of the
charge distribution released by a MIP as seen by this detector side}
\label{fig:figure_4}
\end{center}
\end{figure}

Cutting  fig.~\ref{fig:figure_3} at $\pm 0.5$ and $\pm 1.5$ and shifting the selected parts,
the functions $\{a_{j}(\varepsilon),j=1,\ldots,5\}$ are obtained. The first three $a_j$
are shown in the left side of fig.~\ref{fig:figure_4}, the other two are used in the $x_{g3}$
PDF where five $a_j$ are required.
The accurate elimination
of all the systematic effects, discussed in ref.~\cite{landi03,landi05}, renders
the $a_2(\varepsilon)$ the strip with the maximum energy
in the range $-0.5\leq\varepsilon\leq+0.5$.
Improper corrections give regions
where $a_2(\varepsilon)$ is not the maximum.

The functions $\{a_{j}(\varepsilon),j=1,2,3\}$  allow to
extract the average charge release
by a MIP. The numerical derivative of
the $x_{g3}(\varepsilon)$, built with the  functions $\{a_j(\varepsilon)\}$,
gives an hint of this charge release. The two bumps in the right side of
fig.~\ref{fig:figure_4} are typical of the floating
strip sensors that doubles the MIP shower. It is
noticeable the smoothness and the absence of the
artifacts discussed in ref.~\cite{landi03}.

The histograms of $x_{g2}$ and $x_{g3}$ can be used to test the quality of the functions $\{a_j(\varepsilon)\}$.
The marginal probabilities $H_{x_{g2}}(x,E_t)$ and  $H_{x_{g3}}(x,E_t)$ averaged
over the charge released should coincide with the histograms. The average over the charge released turns
out to be of minor relevance and to
speed up the confronts we use a fixed $E_t$ of 142 ADC counts corresponding to its most probable value.
The  histogram for the $x_{g2}$ data is very similar to the $H_{x_{g2}}(x,E_t)$ ($E_t=142$ ADC), thus for the two strip case
we can assume a good quality of the functions $\{a_j(\varepsilon)\}$. For the three strip case a discrepancy is present
around $x_{g3}\approx 0$. Probably the functions $\{a_j(\varepsilon)\}$ are acceptable even in this case
for the strong sensitivity of
the histograms to these functions. We have to remind the proportionality of the
histograms to $\mathrm{d}\varepsilon/\mathrm{d} x_{g}$.
\begin{figure}[h]
\begin{center}
\includegraphics[scale=0.53]{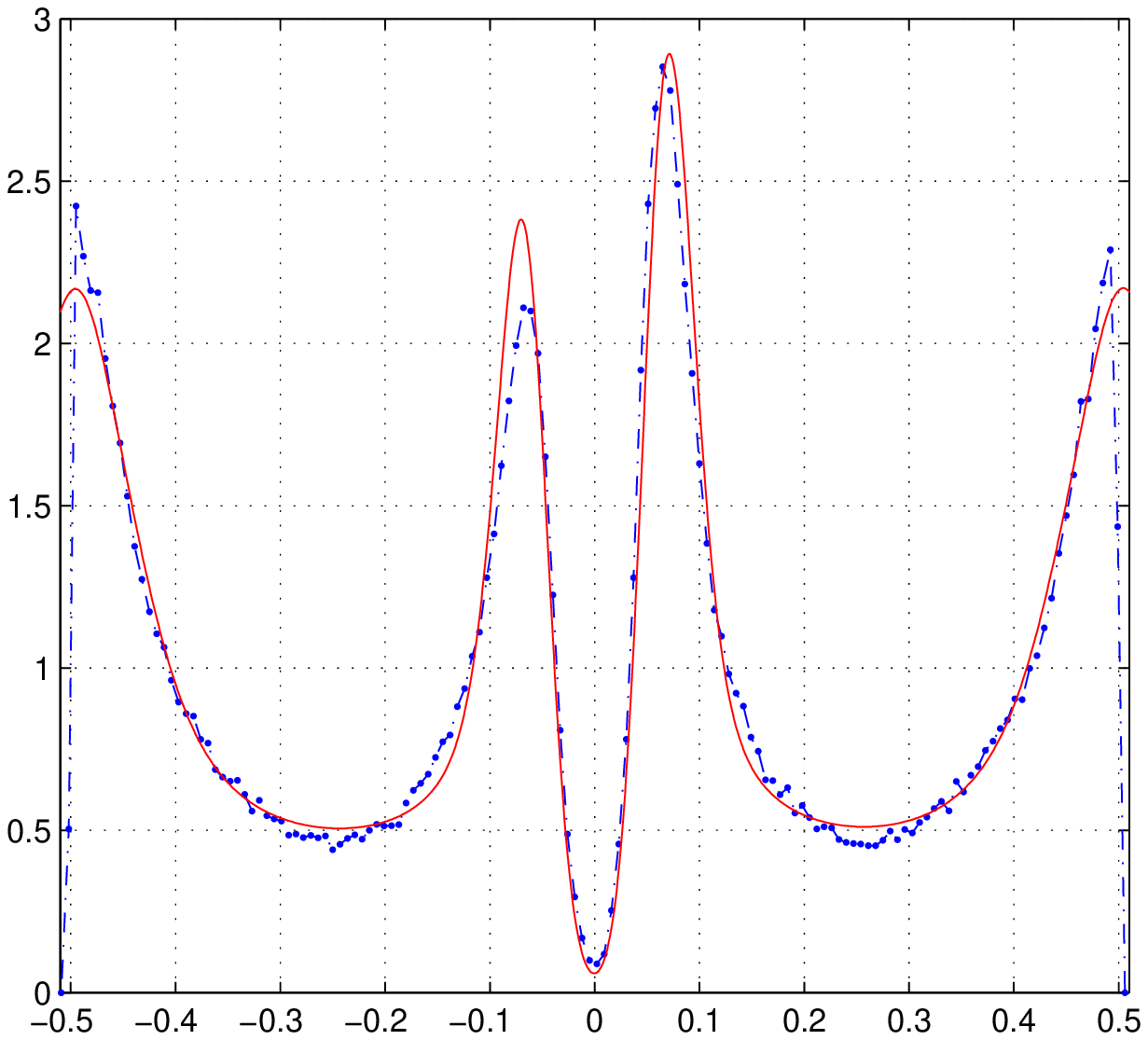}
\includegraphics[scale=0.53]{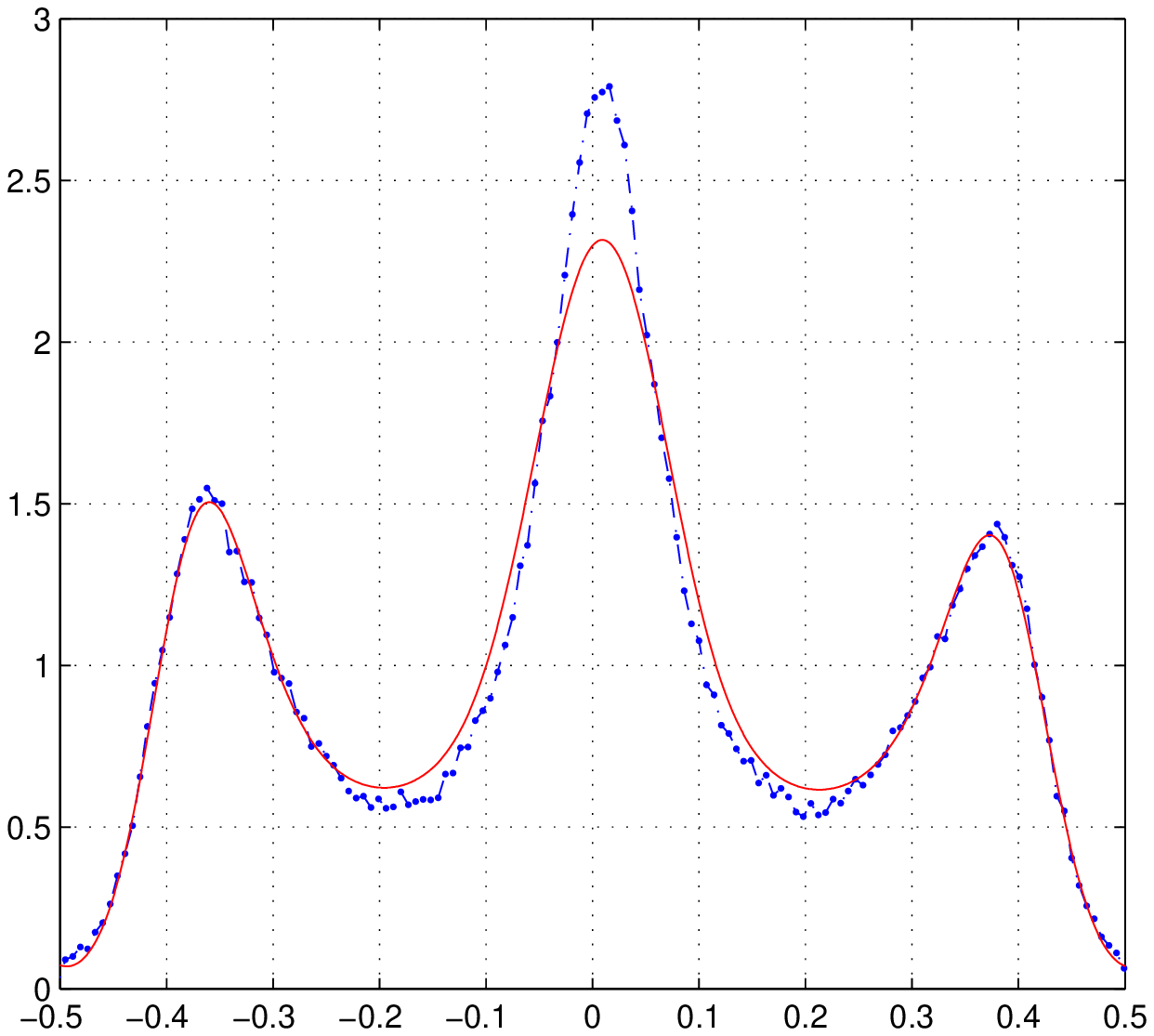}
\caption{ \em To the left, the dotted blue line is the data histogram of $x_{g2}$, the continuous red line is the result of $H_{x_{g2}}(x,E_t)$ and the functions of fig.4 %
with $E_t=142$ ADC-counts. To the right, the same for $x_{g3}$.
 }\label{fig:figure_6}
\end{center}
\end{figure}
\subsection{The simulated data}
The three functions $\{a_j(\varepsilon)\}$ allow the production of
simulated data very similar to the real data,
the following steps synthesize the procedure.
\begin{enumerate}
  \item Random $\varepsilon$-values are generated with a uniform distribution on a strip length ($\tau=1$)
  \item the three $a_j$-values are calculated in each point,
  \item a random number is generated with the distribution of the sum of the three signals
  in the data and used to scale the $a_j$,
  \item The noise: random numbers with gaussian distribution, mean value zero and root mean square 4 ADC counts  are added at each scaled $a_j$,
  \item the simulated data are used  to extract a second generation of $\{a_j\}$-functions to be used in the PDF.
\end{enumerate}
All the generated values are separately saved for future use.
In spite of noise of step $\#3$, the distributions of the simulated data are practically identical to the experimental ones.
The second generation of the $\{a_j\}$-functions turns out to be almost identical to the first one, but, although
the differences are negligible, the calculated histograms have disagreements similar to those of fig.~\ref{fig:figure_6}.
This type of simulation does not allow the introduction of the non uniform charge release along the particle
path. Other simulations, produced with this feature, give results indistinguishable from those without the feature.
The main reason of this insensitivity is probably due to the orthogonal incidence, the diffusion can spread the charge
on the strips in a way very similar to a uniform release. The total charge accounts for the main
part of the fluctuations, the remaining part adds in quadrature to the noise, but, being probably small,
remains invisible. In any case increases of the $\sigma_i$ could absorb these fluctuations.
\subsection{Track reconstruction}
All the simulated tracks are identical: straight lines, incident on the origin and directed
orthogonal to the detector plane. A track is defined by five hits, hence the fit has three degree
of freedom as the PAMELA tracks
in the magnetic field (but the magnetic field was absent in this test beam).

The simulated hits are divided in groups of five to produce the tracks. For the least squares,
the exact impact position $\varepsilon$ of each hit is subtracted from
its reconstructed $\eta_2(x_{g2})$ position, in this way each group of five hits
defines a track with our geometry and the error distribution of $\eta_2(x_{g2})$.
We prefer the use of the $\eta_2(x_{g2})$ positions  because this choice
gives parameter distributions better than those obtained with the simple $x_{g2}$-COG
positions (a frequent choice). The slight improvement is due to the reduction
of the systematic errors present in the COG.
In the $\{{\xi},{z}\}$-plane the tracks have equation:
\begin{equation}\label{eq:equation_13a}
    \xi=\gamma {z}+\beta.
\end{equation}
By definition, all the tracks have $\gamma=0,\,\beta=0$, but, due to the noise,
their fitted values are distributed around zero. For our PDF, no linear
reduction is possible and we have
to handle the non-linearity of the likelihood maximization. Therefore,
the parameters $\{\widetilde{\gamma_n},\widetilde{\beta_n}\}$ of the track $n$ are obtained minimizing $L(\gamma_n,\beta_n)$
defined as the negative logarithm of the likelihood with the PDF of eq.~\ref{eq:equation_13}:
\begin{equation}\label{eq:equation_14}
    L(\gamma_n,\beta_n)=\Big(-\sum_{j=5n+1}^{5n+5}\,\ln[P_{x_{g2}}(x(j),E_t(j),\gamma_n {z}_j+\beta_n+\varepsilon(j))]\Big)
\end{equation}
$x(j)$, $E_t(j)$ and ${z}_j$ are respectively: the $x_{g2}$-value of
the two strip COG, the total signal of the hit $j$ for the track $n$ and the position of the $j-5n$ detector plane.
The functional dependence $\gamma_n {z}_j+\beta_n+\varepsilon(j)$ is inserted
in the $\varepsilon$-dependence of
$\{a_k(\varepsilon)\,k=1,2,3\}$, $\varepsilon(j)$ must be added to
recover the shift to have impact points on tracks with $\{\gamma=0,\beta=0\}$.

\begin{figure}[b]
\begin{center}
\includegraphics[scale=0.55]{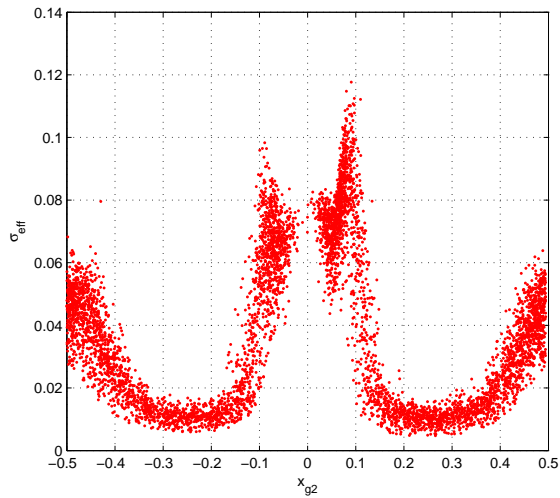}
\caption{\em  Distribution of the effective standard deviations $\{\sigma_{eff}(j)\}$ in function of their $x_{g2}(j)$ positions.
 }\label{fig:figure_7}
\end{center}
\end{figure}
The minimizing algorithm is the standard
$\mathrm{MATLAB}$~\cite{matlab} $\mathit{fminsearch}$ function. As a starting point
of the minimum search, we could use the parameters given by the least squares results.
But, their precisions are modest and
it would be better to have a nearer starting point. For this we try to approximate the
probability distributions at fixed
$x_{g2}$ with gaussians.  For each hit, we calculate an effective variance ($\{\sigma_{eff}(j)^2\}$)
and use it as the width of a supposed gaussian error. The range of integrals for
$P_{x_{g2}}(x,E_t,\varepsilon)\varepsilon^2$
must be drastically limited to avoid the divergence of the
Cauchy-like tails. The effective gaussian of each hit is
centered in the $\eta_2(j)-\varepsilon(j)$ with a width $\sigma_{eff}(j)$.

The distribution of the effective standard deviation
($\{\sigma_{eff}(j)\}$) shows appreciable variations
along the strip (fig.~\ref{fig:figure_7}),
the scale of this figure amplifies the variations.
Much larger variations will be encountered in the following.
The trend of the $x_{g2}$-histogram is easily recognizable
in the $\{\sigma_{eff}(j)\}$ distribution.
This is due to a relation between these two plots.
In facts, $\{\sigma_{eff}(j)\}$ estimates
the range of possible $\varepsilon$-values corresponding
to an $x_{g2}$-value, but, similarly for the histograms,
the height of the $x_{g2}$-value is given by the $\varepsilon$-interval
that produces the same $x_{g2}$.
Thus,  the highest $\{\sigma_{eff}(j)\}$
are located in the highest regions of the histogram, and the lowest
$\{\sigma_{eff}(j)\}$ are in the lowest regions of the histogram.
Here the assumption of uniform "illumination" is again essential.

The effective gaussian approximations reduce the maximum likelihood search
to linear equations, their solutions
are used as starting points for the $\mathrm{MATLAB}$ $\mathit{fminsearch}$
function to minimize eq.~\ref{eq:equation_14}.
{\em In the following, we will call MIN-LOG the parameters
$\{\widetilde{\gamma_n},\widetilde{\beta_n}\}$ given by the minima
of eq.~\ref{eq:equation_14}.
With $"$effective variance$"$ we will indicate
the results obtained with the effective gaussian parameters
$\{\sigma_{eff}(j),\eta_2(j)-\varepsilon(j)\}$ (weighted least squares) for each hit} and
these two methods will be compared with the results of a
least squares approach, often the baseline of track fitting. (The last two approaches
are often called $\chi^2$ minimization.)
The Kalman filter is not essentially different from least squares,
it has important advantages for its recursiveness in complex environments.

\begin{figure}[b]
\begin{center}
\includegraphics[scale=0.53]{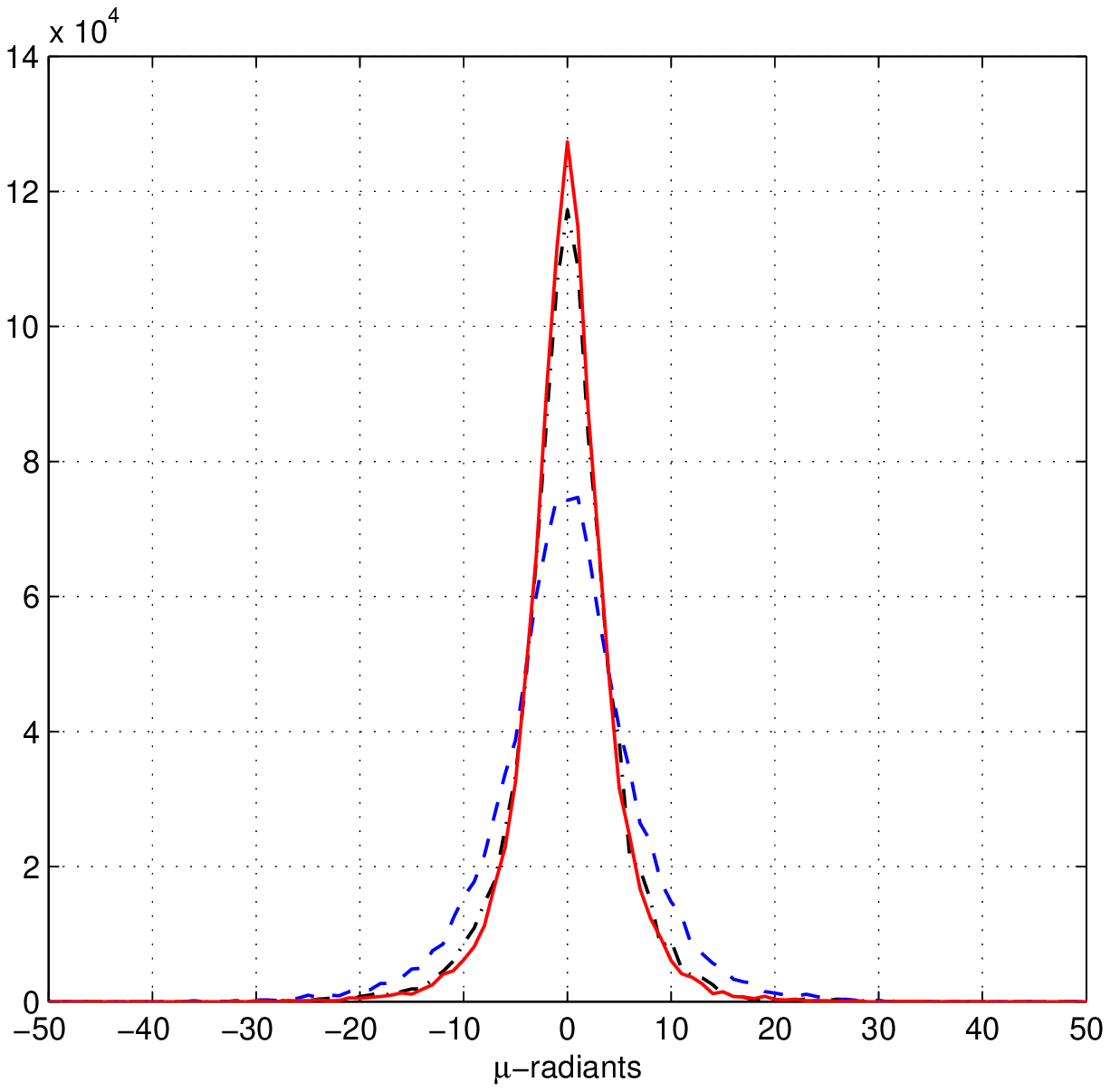}
\includegraphics[scale=0.53]{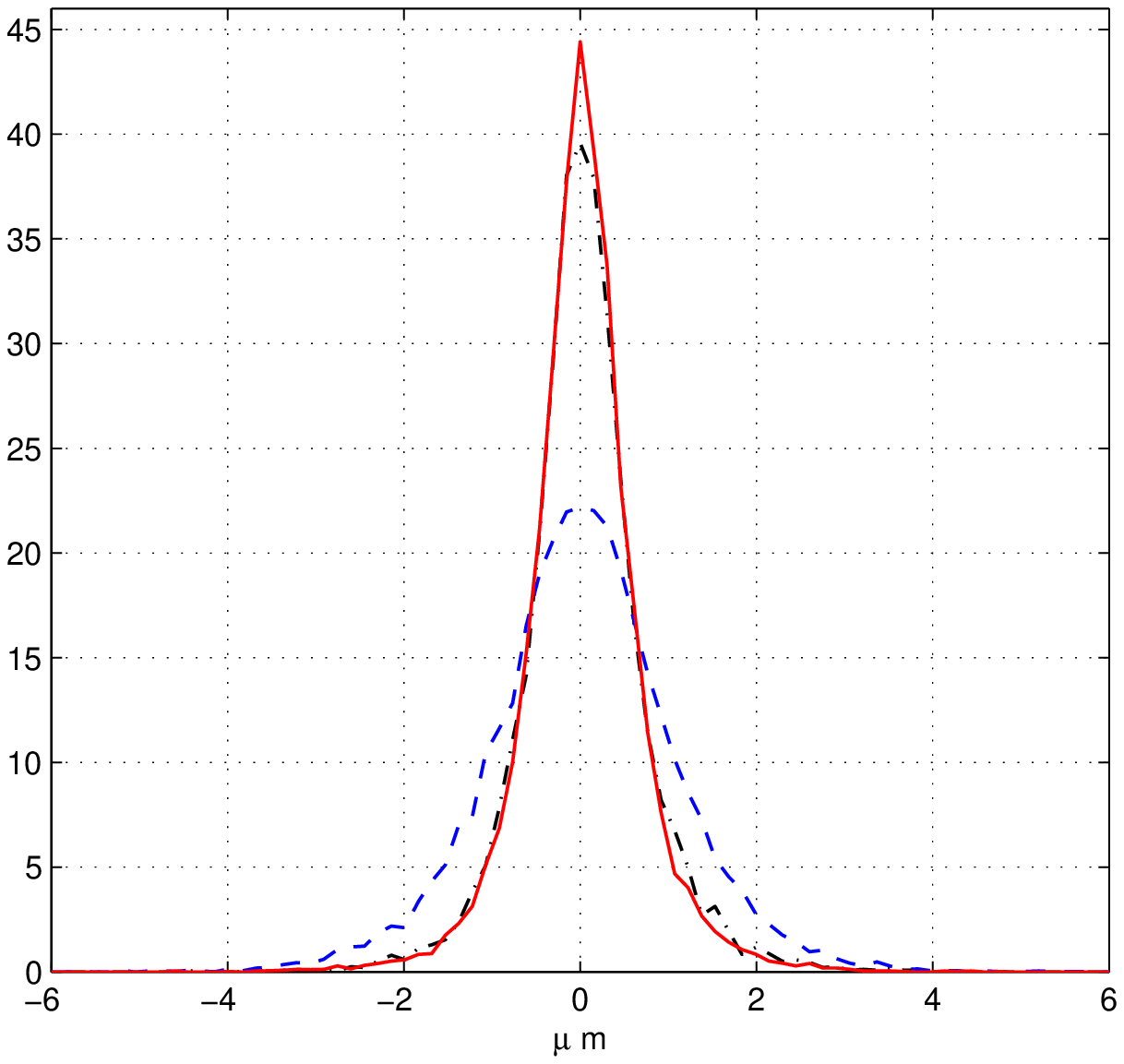}
\caption{\em Floating strip side. Distributions of the track parameters $\gamma$ (left) and $\beta$ (right). Continuous red lines: MIN-LOG. Dash dotted black lines: effective variances.  Dashed blue lines: least squares.
 }\label{fig:figure_8}
\end{center}
\end{figure}
We can see in fig.~\ref{fig:figure_8} the drastic
improvements of these two methods: the MIN-LOG
and the use of the effective variances
respect to the least squares method.
The extraction of the FWHM is an annoying
procedure, but we have no other way to compare distributions
with large tails. The ratios of the
FWHM of the least squares and the MIN-LOG are around a factor
two. The detailed values are: the ratio
of the FWHM for $\gamma$ is 1.7, that for $\beta$ is 2.17,
the ratio of the maxima (MIN-LOG divided by least squares)
is 2.3 for $\gamma$ and 2.0 for $\beta$.
The MIN-LOG has a slightly better distributions of $\gamma$ or
$\beta$ than the that obtained with the effective
variances. The strong similarity of the results are due
to the good approximation with  gaussians of our PDF, here the
differences of the tails are irrelevant.  The drastic
simplification of the linear equations suggests
this method as a viable alternative to minimization
of eq.~\ref{eq:equation_14}, even if the computational
complexity is comparable. The
extraction of $\sigma_{eff}(j)$ is a time consuming procedure.

Our preference to explore the distributions of the $\gamma$
and $\beta$ parameters is due to the acceptable similarity with the
least squares results and we are directly interested in them as the aim
of the fit.
Another alternative would be the
exploration of the residuals, their extensive use is due to
easy extractions from the data and the assumption
of a direct relation with the error PDF. Now the process is very complex
and the residual distributions
turn out to be extremely different from these given by the least squares,
our PDF produces very high peaks around zero
due to the contacts of the tracks with good hits. Even
if accessible from the data, the residuals do not show a clear connection
to our preferred plots of fig.~\ref{fig:figure_8}.

Another type of interesting residuals are the differences
respect to the exact points. These are easily produced by the simulations,
but, showing relations similar to those of fig.~\ref{fig:figure_8},
they are discussed in the last subsection.

\begin{figure}[h]
\begin{center}
\includegraphics[scale=0.53]{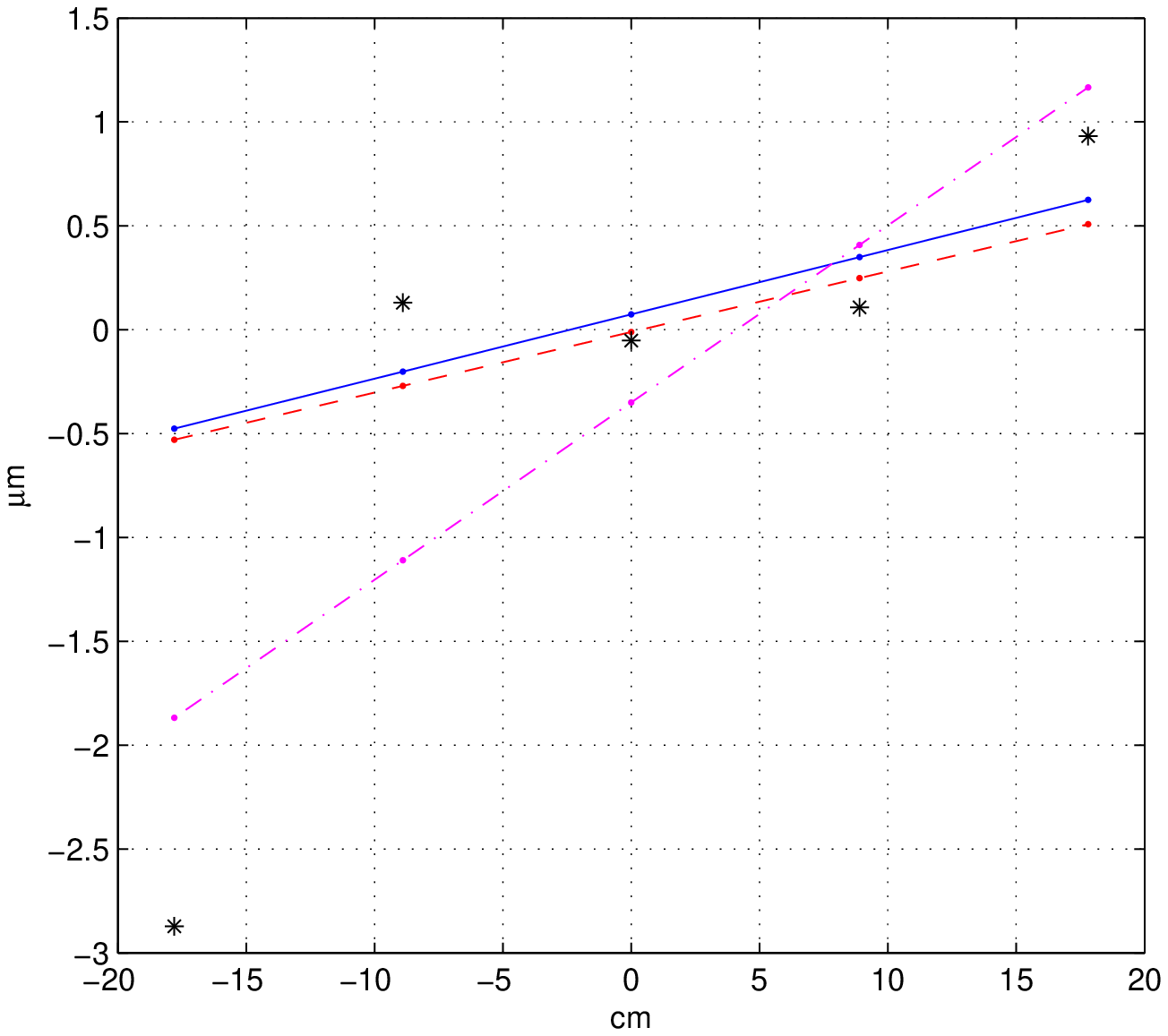}
\includegraphics[scale=0.53]{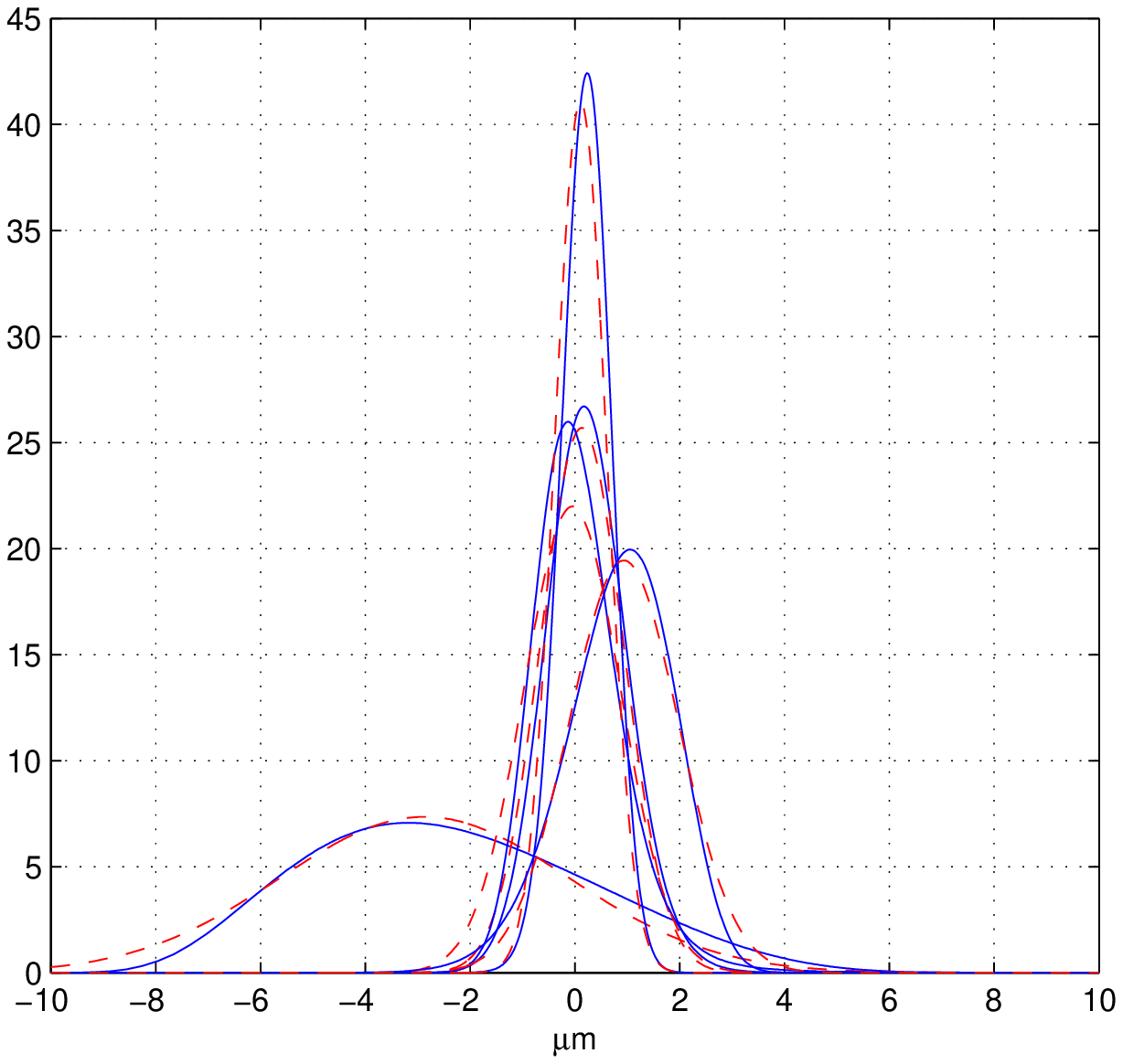}
\caption{\em To the left, the reconstructed track with the parameters given by the three approach. Continuous
blue line: parameters from MIN-LOG.
Dashed red line: parameters from "effective variance". Dash dotted magenta line: least squares.
To the right,  the continuous blue lines are the $P_{x_{g2}}(x(j),E_t(j),\varepsilon-\varepsilon(j))$
for the hits of this track.
Dashed red lines are their approximate gaussian PDF with
variances $\sigma_{eff}(j)$ and centered in $\eta_2(j)-\varepsilon(j)$.
 }\label{fig:figure_9}
\end{center}
\end{figure}
The agreement of the effective gaussian approximations to our
PDF is illustrated in fig.~\ref{fig:figure_9},
the two types of PDF are plotted together
for the five hits of a track. For each hit, we observe
different distributions with the gaussian approximations
very similar to the our PDF. The third hit is very good,
with the narrowest probability distribution, and the
reconstructed track passes near to it. The first hit is almost discarded
having a wide probability distribution and a position evidently out of line.
The other hits contribute to the
slight bending of the track. We have to remind the differences
of the tails of the PDF, invisible at this scale,
they do not play a role on this sensor side, but are relevant in the other side.

To verify the consistency of our
PDF, we can observe in fig.~\ref{fig:figure_9} the strict proximity of the maxima of the
peaks for the narrow distributions.
We have to recall that the sole element obtained from  $\{P_{x_{g2}}(x(j),E_t(j),\varepsilon)\}$
are the $\{\sigma_{eff}(j)\}$ of the effective gaussian PDF, and
integrals on $\{P_{x_{g2}}(x(j),E_t(j),\varepsilon)\}\varepsilon^2$ give their widths.
The center of each gaussian is
$\eta_2(j)-\varepsilon(j)$ imposed by our virtual tracks $\{\gamma=0,\beta=0\}$.
The maximum of the peak for $P_{x_{g2}}(x(j),E_t(j),\varepsilon)$
(shifted of the identical $\varepsilon(j)$ )
is produced by the three functions $a_j(\varepsilon)$ inserted
in the PDF. The functions $a_j(\varepsilon)$ are obtained with a set of
transformations (from eq.~\ref{eq:equation_3} to eq.~\ref{eq:equation_12}) over the
averages of eq.~\ref{eq:equation_12a}, and the sliding window covers a large set of
charges deposited in their corresponding positions $\eta_j(x_{gj})$. These
two completely different paths give very consistent results. At
the same time, fig.~\ref{fig:figure_9} underlines the inconsistency of using
directly the COG in the fit. In this case, the effective (or constant) gaussian
acquires a variable shift respect to its corresponding
$P_{x_{g2}}(x(j),E_t(j),\varepsilon)$ distribution,
and part of the COG systematic error enters the fit.

\section{Normal strip side}

In the normal strip side, all the strips are connected to the read out system.
The charge spread is lower and the noise
is higher than in the floating strip side. The strip noise of $8$ ADC
counts produces large shifts of the reconstructed points,
and the long ranges of the  $P_{x_{g2}}(x,E_t,\varepsilon)$ turn out to be relevant.
The function $f(-\varepsilon)$  is similar to an interval function, the
slight rounding to the sides is due to a small charge
diffusion on the adjacent strips. The large noise increases the
reconstruction artifacts around $\varepsilon\approx\pm 1$,
and around $\varepsilon\approx 0$.

\begin{figure}[h]
\begin{center}
\includegraphics[scale=0.53]{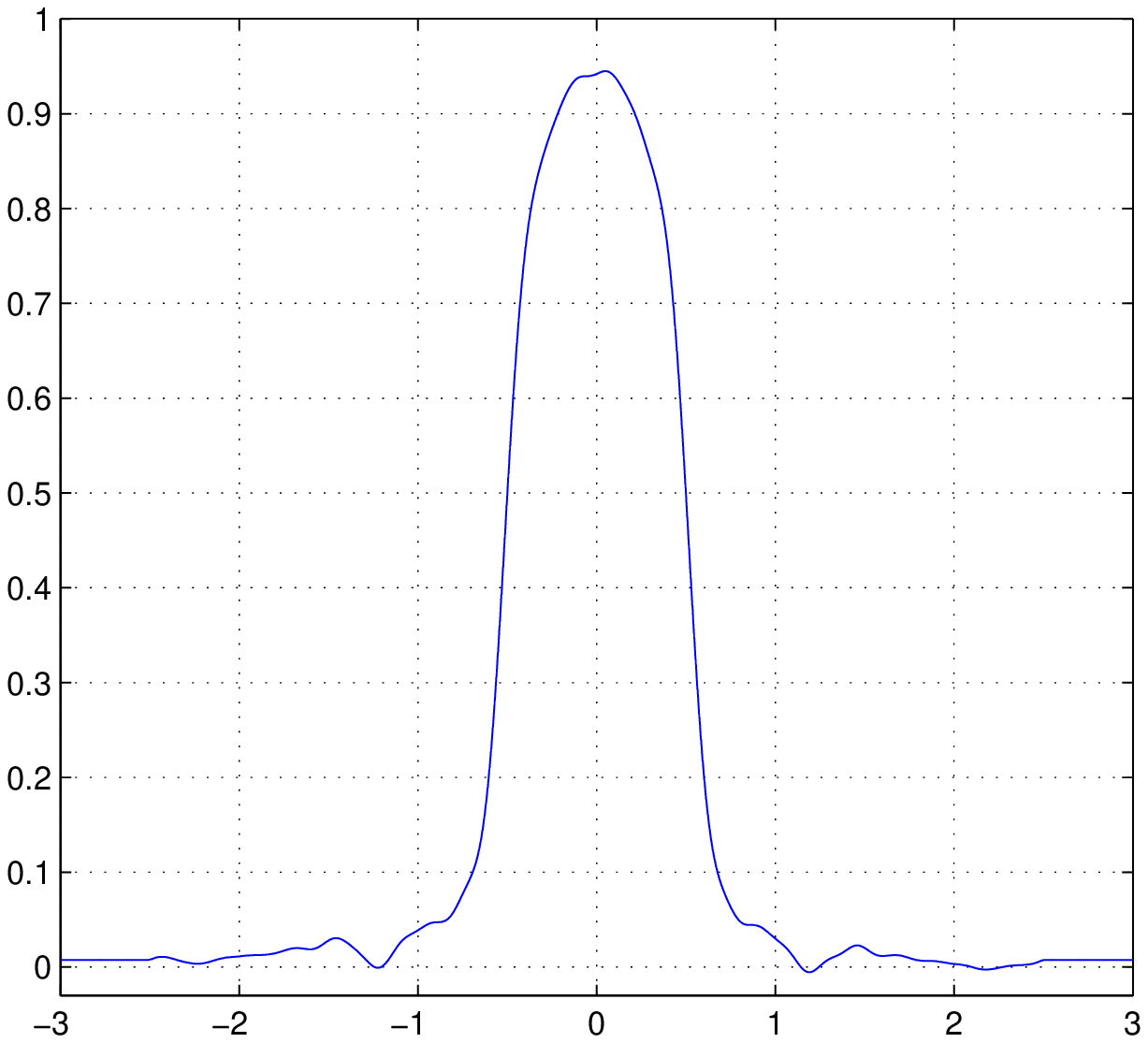}
\includegraphics[scale=0.53]{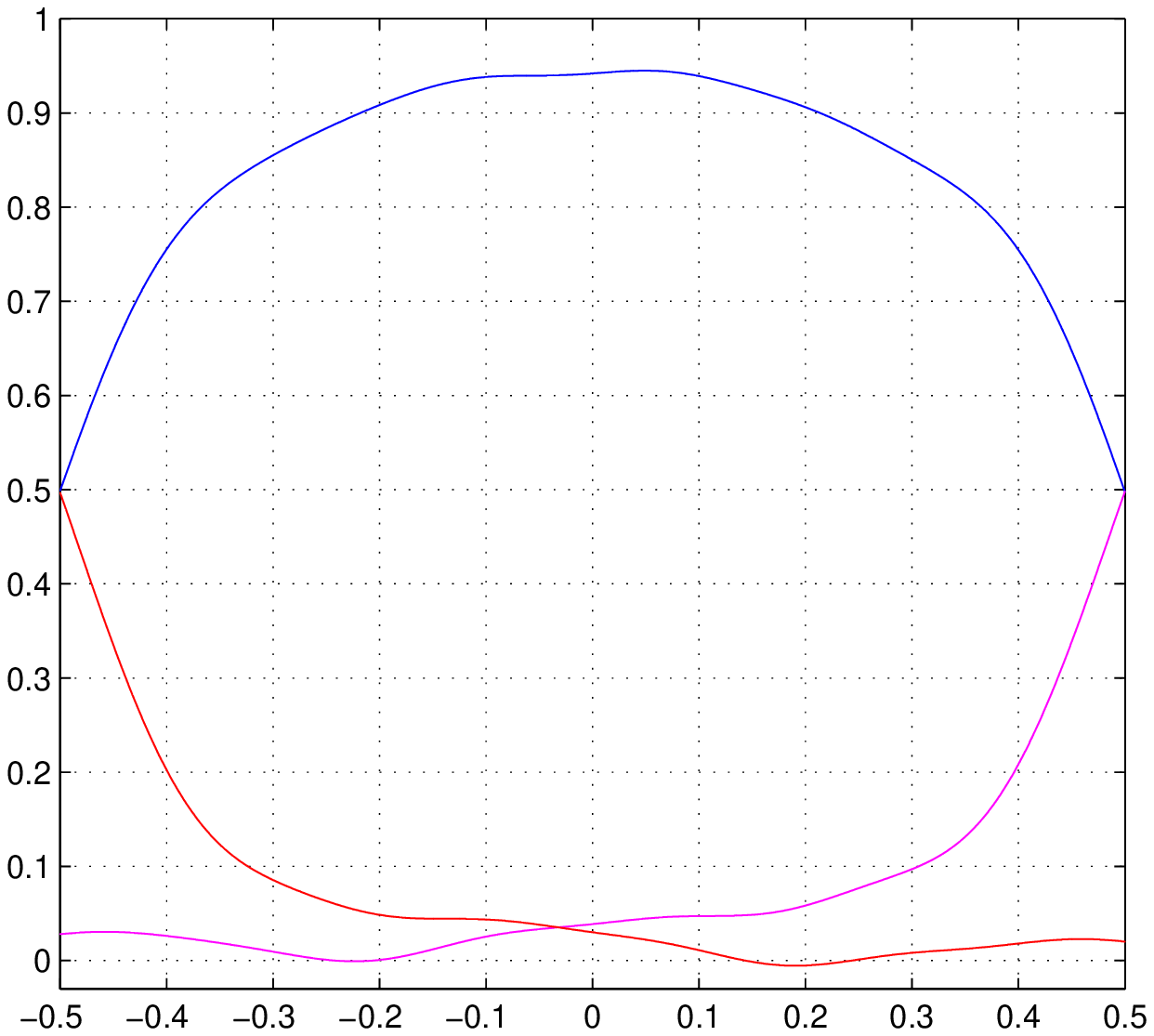}
\caption{\em  Normal strip side. To the left the function $f(-\varepsilon)$ reconstructed from the data of this side. To the right the three energies $a1(\varepsilon)$ (magenta line), $a2(\varepsilon)$ (blue line), $a3(\varepsilon)$ (red line) for this side .
 }\label{fig:figure_10}
\end{center}
\end{figure}
\begin{figure}[t]
\begin{center}
\includegraphics[scale=0.53]{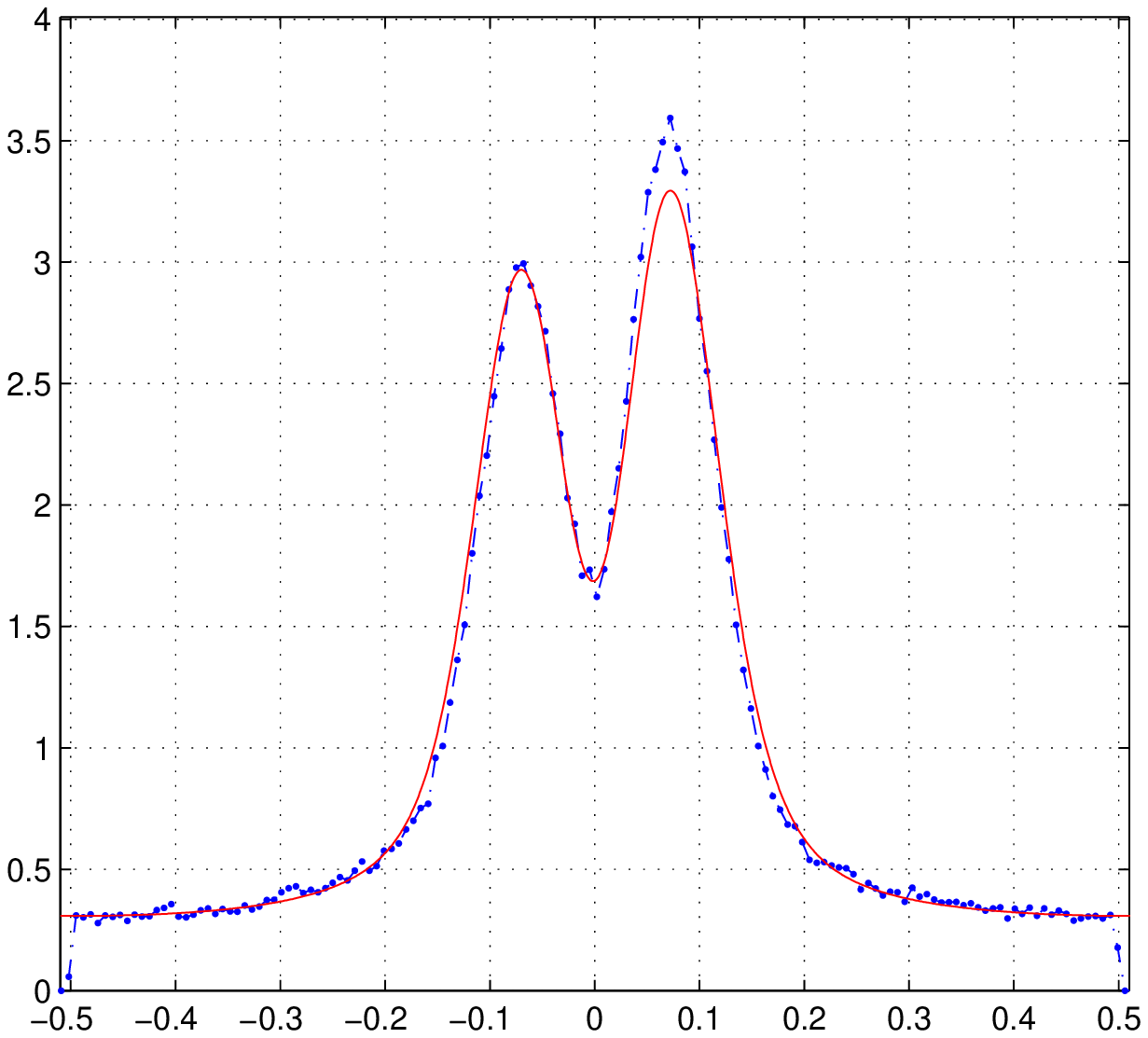}
\includegraphics[scale=0.53]{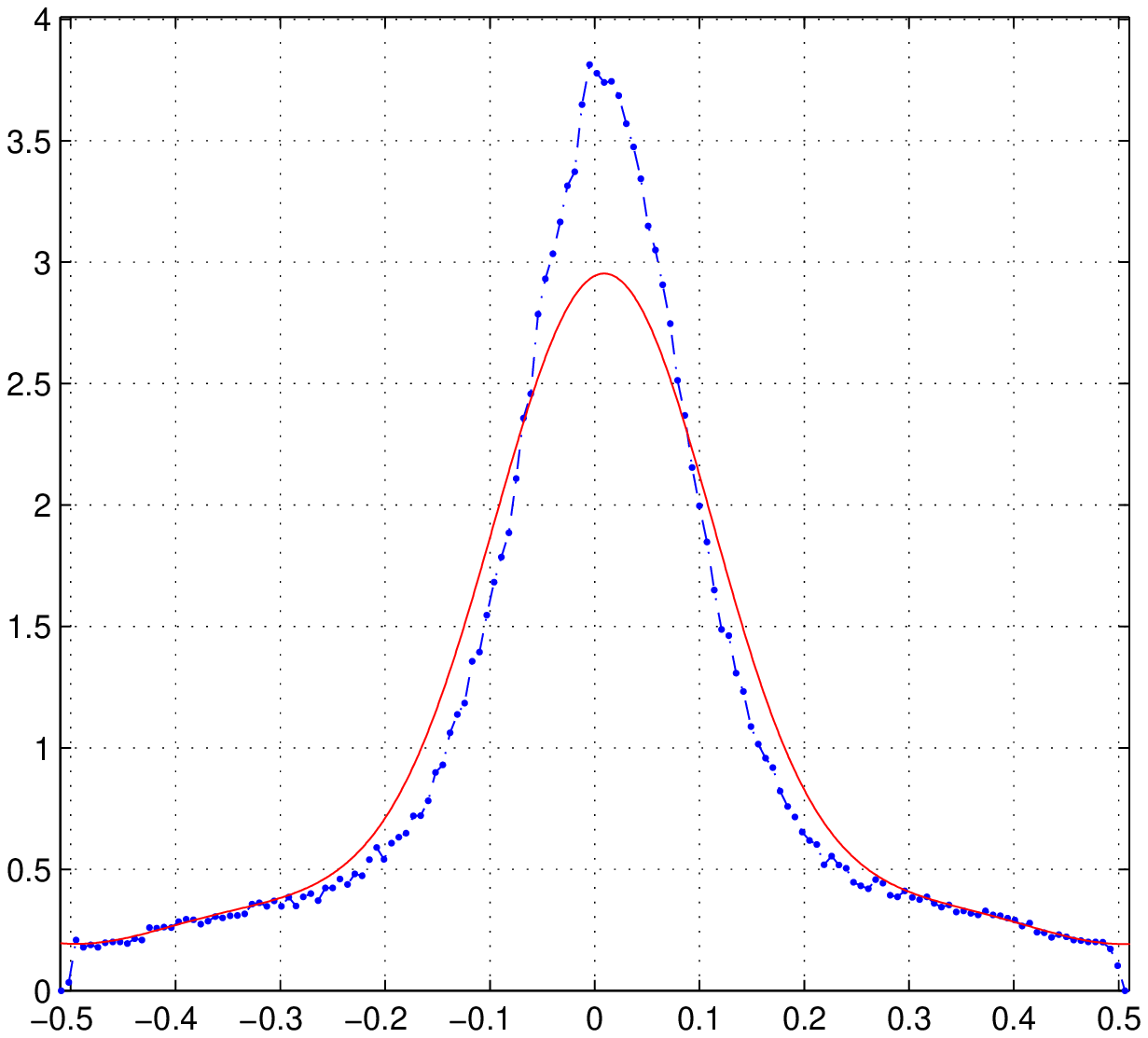}
\caption{\em  To the left the $x_{g2}$ histogram, the red line is our calculated probability distribution
at a fixed energy (142 ADC). To the right the $x_{g3}$ histogram and the probability distribution at a fixed energy.
}\label{fig:figure_11}
\end{center}
\end{figure}

The forms of the $\{a_j(\varepsilon)\}$ modifie appreciably the reproduction of $x_{g3}$ histogram,
but they have a negligible effect on the
$x_{g2}$ histogram as illustrated in fig.~\ref{fig:figure_11}. The good consistency with the $x_{g2}$
histogram is an indication
that the $\{a_j(\varepsilon)\}$ could be well suited to an application
using two strips at time.

Even in this case, the  energy fractions $\{a_j(\varepsilon)\}$ are
used to produce simulated events with the steps of section
$4.1$. Now, the large noise forces us to modify point $3)$.
The distribution of the scaling factors  must be
cleaned from the noise, and the agreement to the data can be
achieved after the addition of random noise (8 ADC r.m.s.) at
each signal of any strip.
As for the floating strip side, the simulated hits are collected in
groups of five,
and a set of virtual straight tracks, with  $\gamma=0$ and
$\beta=0$ in eq.~\ref{eq:equation_13a}, is obtained after
subtracting the impact point of each hit. The three degree of
freedom for this side are less than those of the PAMELA tracker,
but the five hit tracks allow a comparison
with the floating strip side.

\begin{figure}[t]
\begin{center}
\includegraphics[scale=0.53]{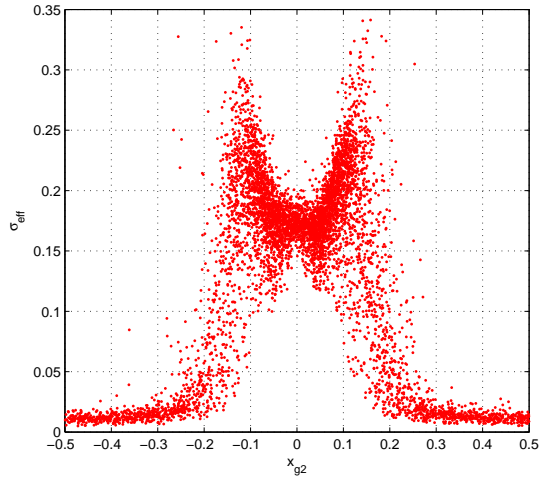}
\caption{\em  Distribution of effective standard deviations $\sigma_{eff}(j)$ for the normal strip side
}\label{fig:figure_12}
\end{center}
\end{figure}

Equation~\ref{eq:equation_14} is used to reconstruct the
track parameters. The minimum search algorithm
is initialized as above extracting
an effective variance $\sigma_{eff}(j)^2$. Even here,
our PDF for each hit must be cut to capture the main
part where the probability is higher and to avoid the
divergences given by the tails. But now, the approximations of our PDF with gaussians are often poor.
The distribution of the parameters
$\{\sigma_{eff}(j)\}$ (fig.~\ref{fig:figure_12}) has
a large anisotropy along the strip, and again shows a strong
similarity with the $x_{g2}$ histogram.
Hence the rule of fig.~\ref{fig:figure_7} is respected, the highest regions of
the histogram are connected to the highest
$\{\sigma_{eff}(j)\}$ and the lowest
regions of the histogram are connected to the lowest
$\{\sigma_{eff}(j)\}$.
Below $|x_{g2}|\approx 0.3$ the hits have very
large $\sigma_{eff}(j)$'s. For the rest of the range of
$x_{g2}$, the hits have $\sigma_{eff}$
comparable with those of the floating strip side.

\subsection{Reconstruction details}

\begin{figure}[b]
\begin{center}
\includegraphics[scale=0.5]{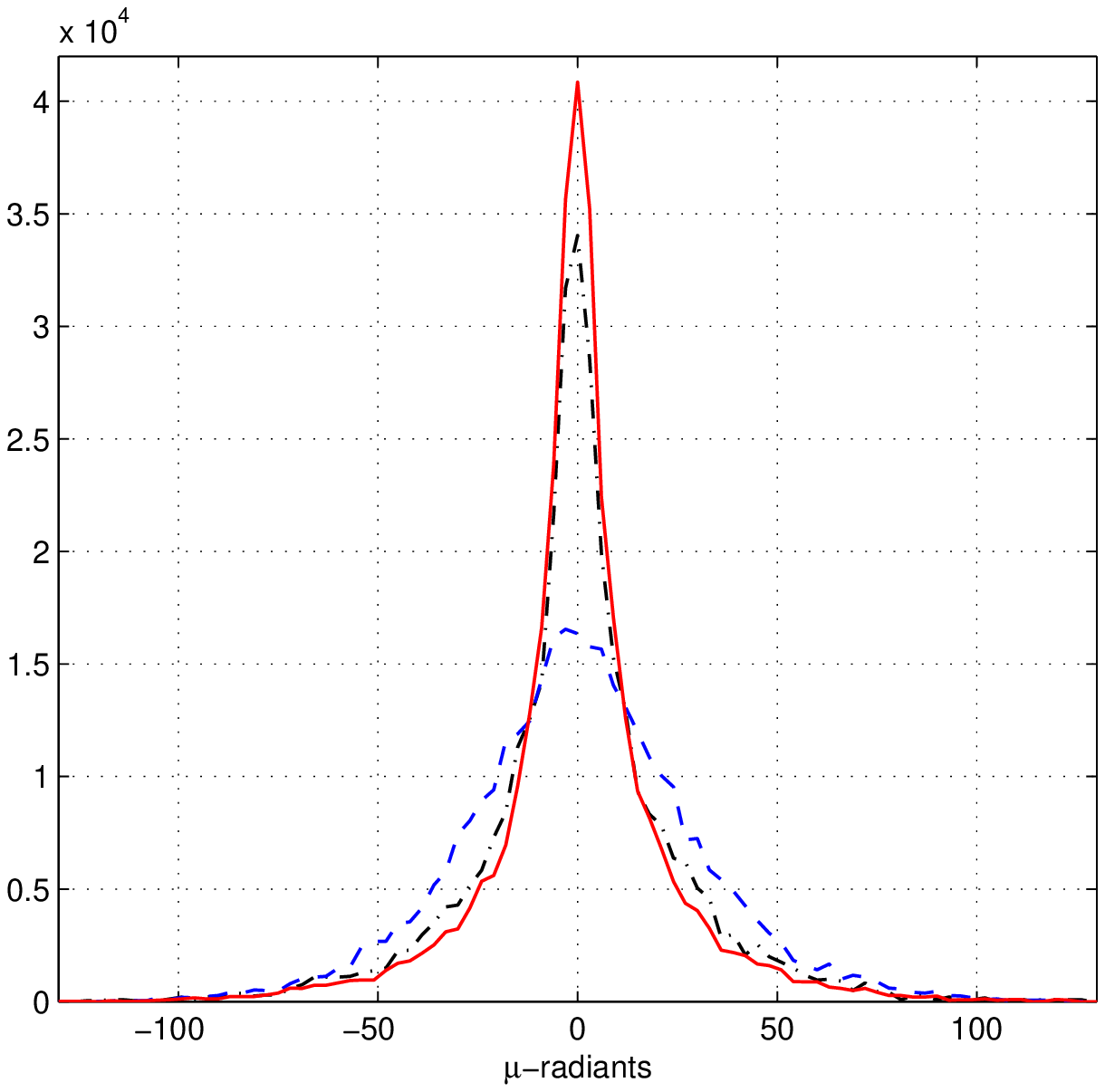}
\includegraphics[scale=0.5]{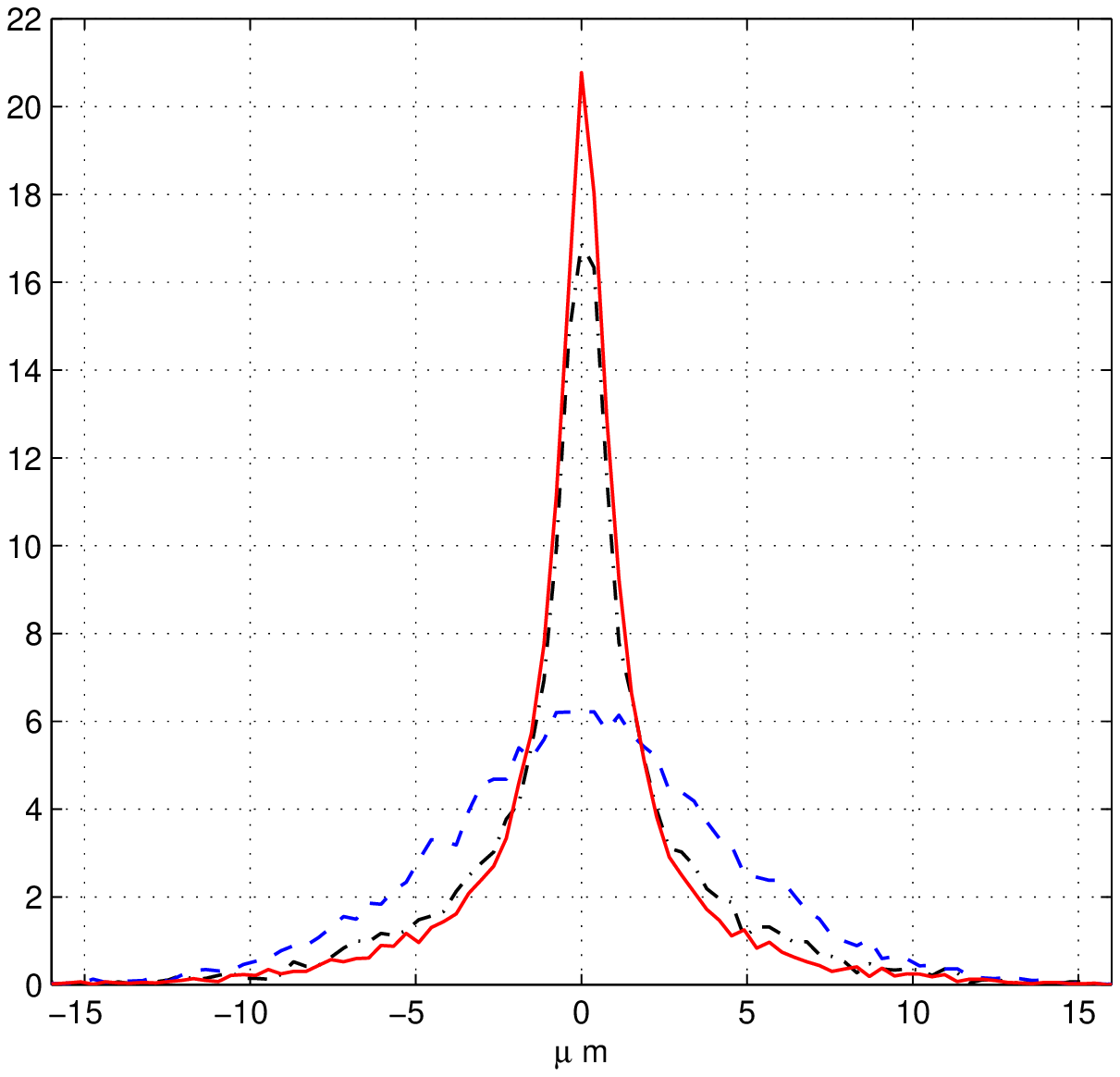}
\caption{\em Normal strip side. Distribution of $\gamma$ (left)
and $\beta$ (right). Continuous red line: the MIN-LOG, dashed blue line:
the least squares, dash dotted black line: the effective variance.
}\label{fig:figure_13}
\end{center}
\end{figure}

The improvement  given by the minima of eq.~\ref{eq:equation_14}
respect to those of the least squares is appreciably larger than that of the floating strip side.
The FWHM of the parameter distributions are
better by a factor 3 respect to the least
squares approach as shown in fig.~\ref{fig:figure_13}. The precise values
of the ratios for the FWHM are: 3.6 for the
$\gamma$ distributions and 4.8 for the $\beta$ distributions.
The ratio of the peak values are 2.6 for the
$\gamma$ distributions and 3.4 for the $\beta$ distributions.
We assume a conservative factor 3 for the global improvement
of the minima of eq.~\ref{eq:equation_14} (MIN-LOG) respect to the least squares method.
The effective variances give parameter distributions a
little wider than those obtained by eq.~\ref{eq:equation_14}, but always much
better than the least squares.

To understand the origin of these large improvements,
we have to look to the hits and their reconstructed tracks,
trying to recover some rules. An immediate explanation
can be given by the observation of the $\sigma_{eff}$ distribution
reported in fig~.\ref{fig:figure_12}.
If two hits in a track are in the regions with low $\sigma_{eff}$, they
define almost completely the track parameters. The least squares
method has not this information, all the events are equally important
and the noise displays its full effect.
\begin{figure}[h]
\begin{center}
\includegraphics[scale=0.43]{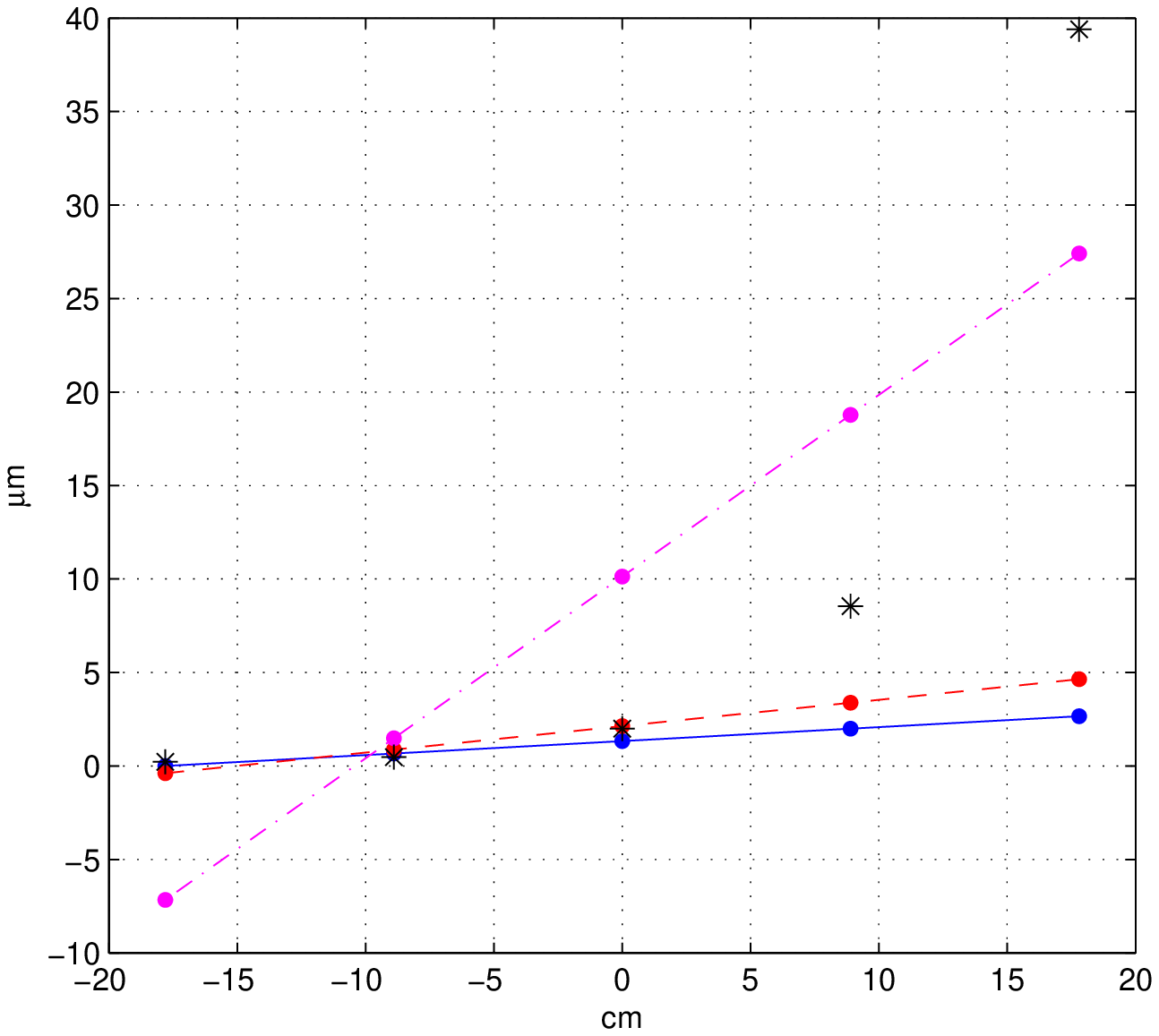}
\includegraphics[scale=0.45]{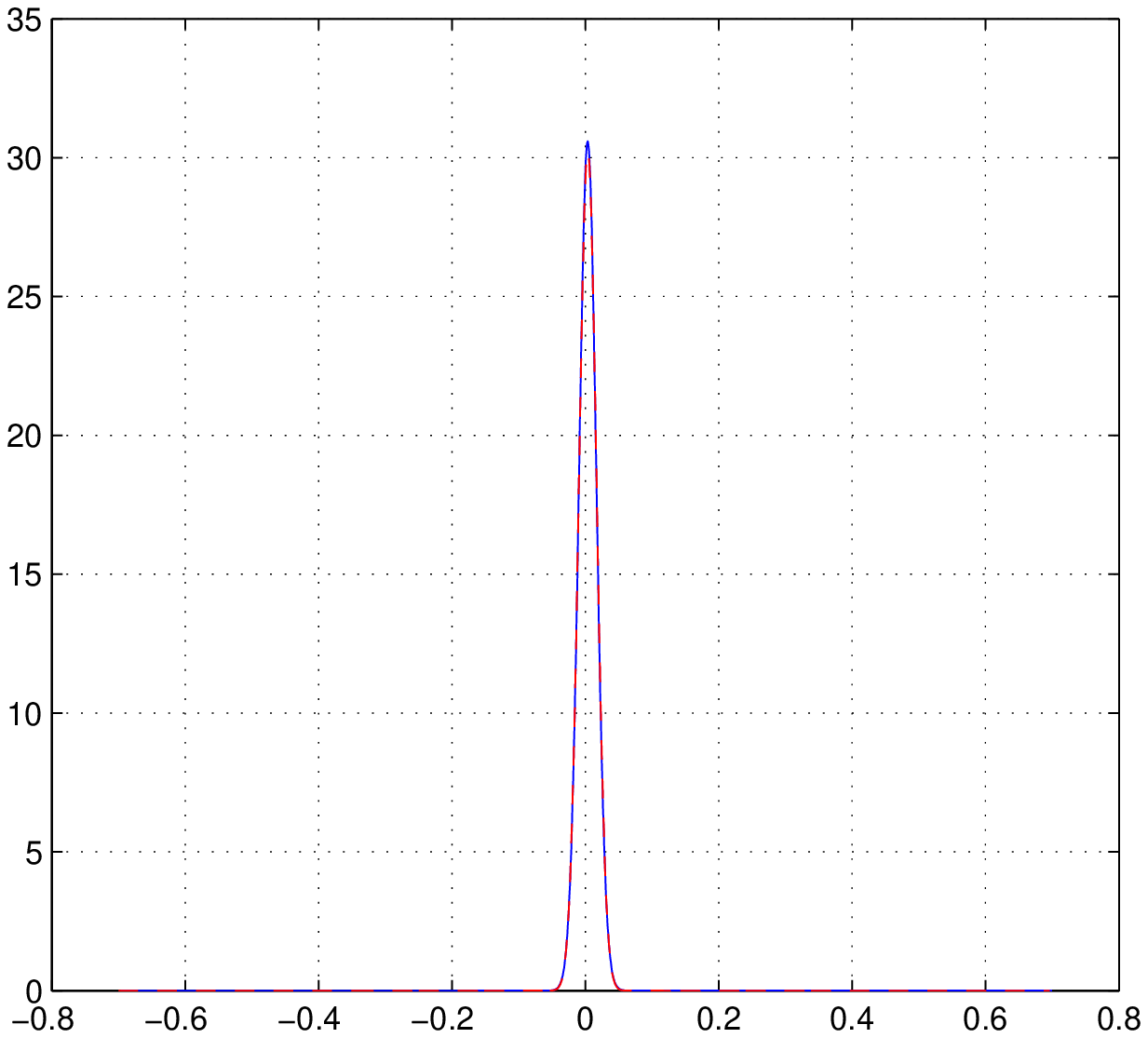}
\includegraphics[scale=0.45]{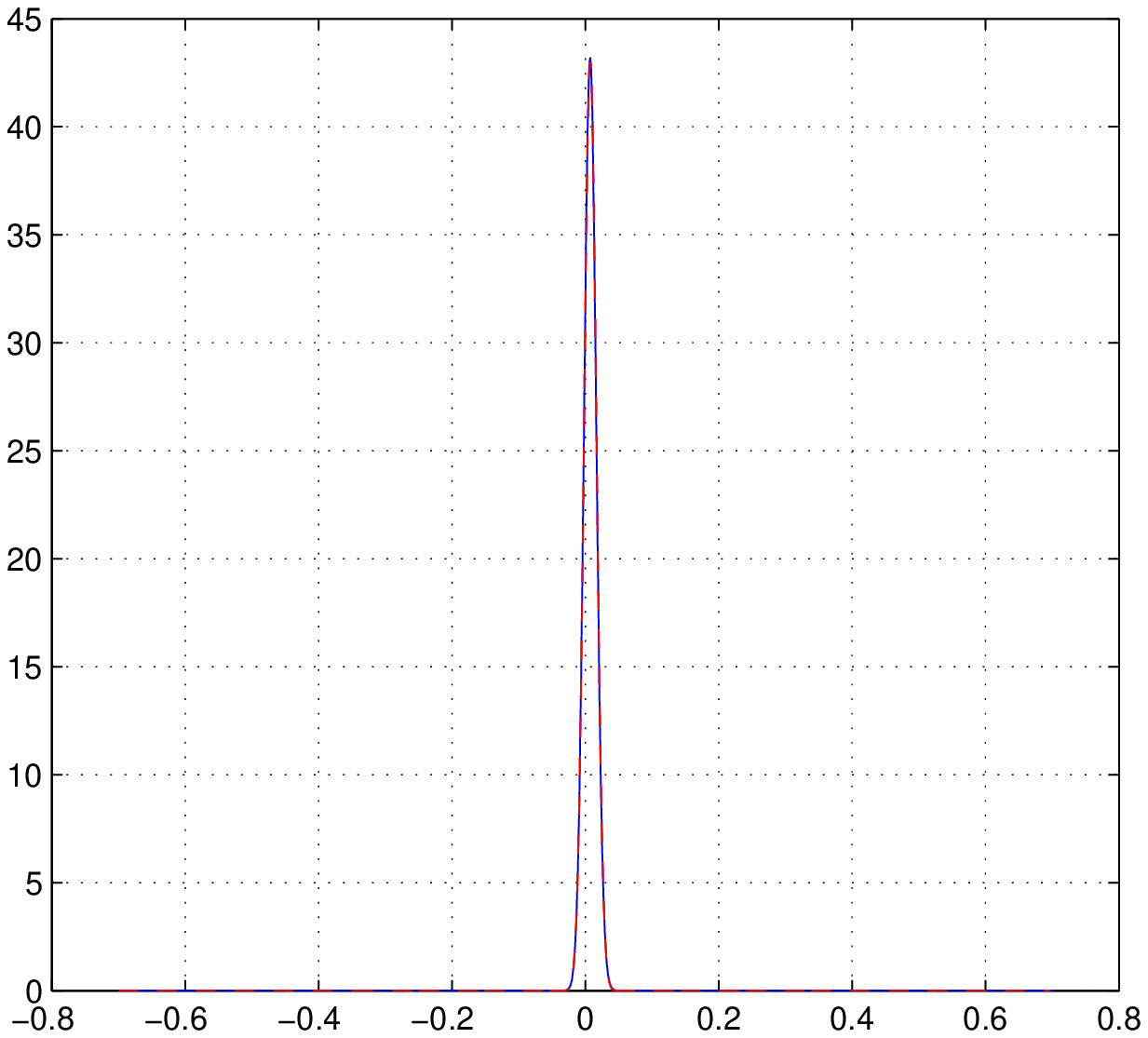}
\includegraphics[scale=0.45]{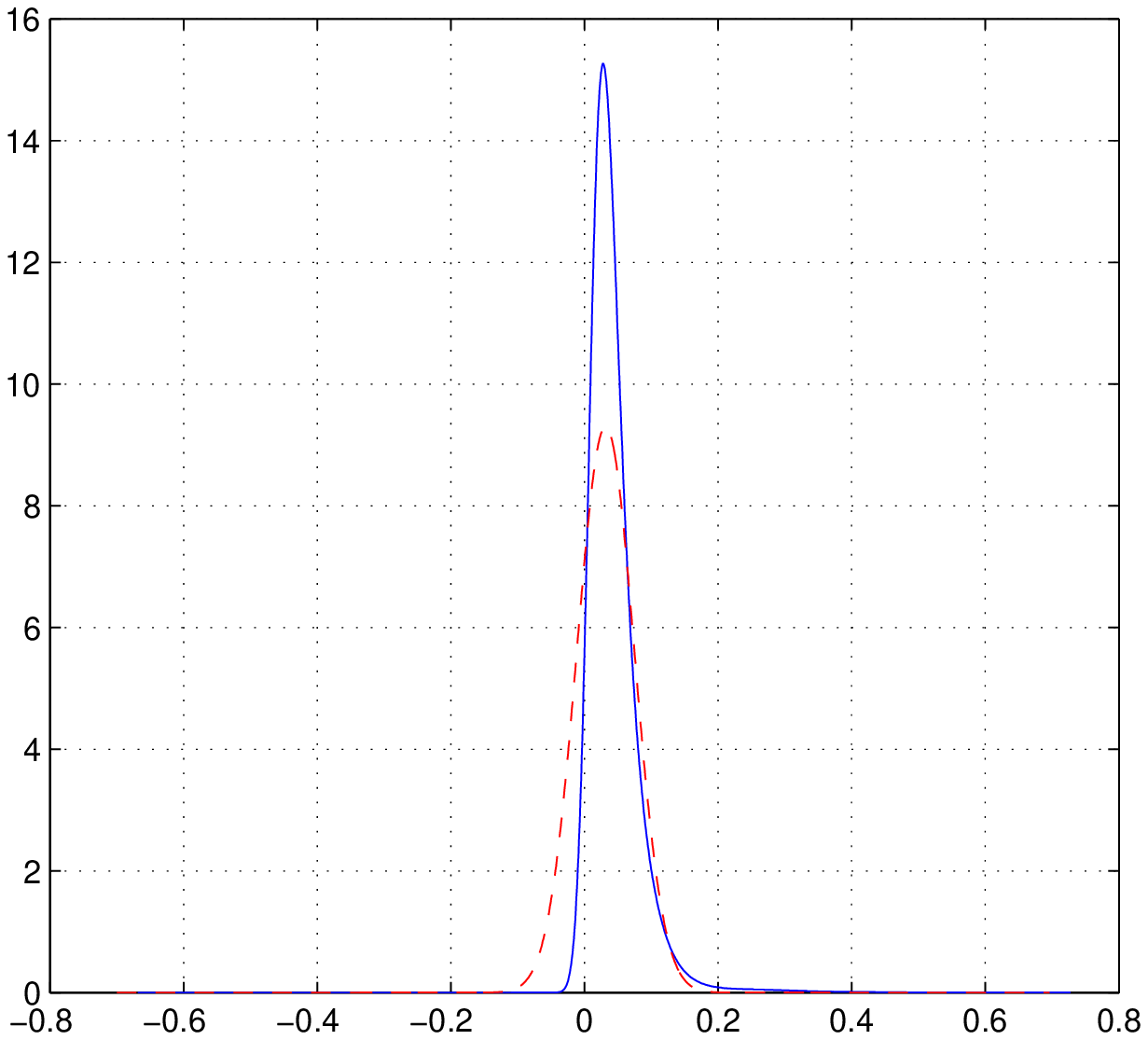}
\includegraphics[scale=0.45]{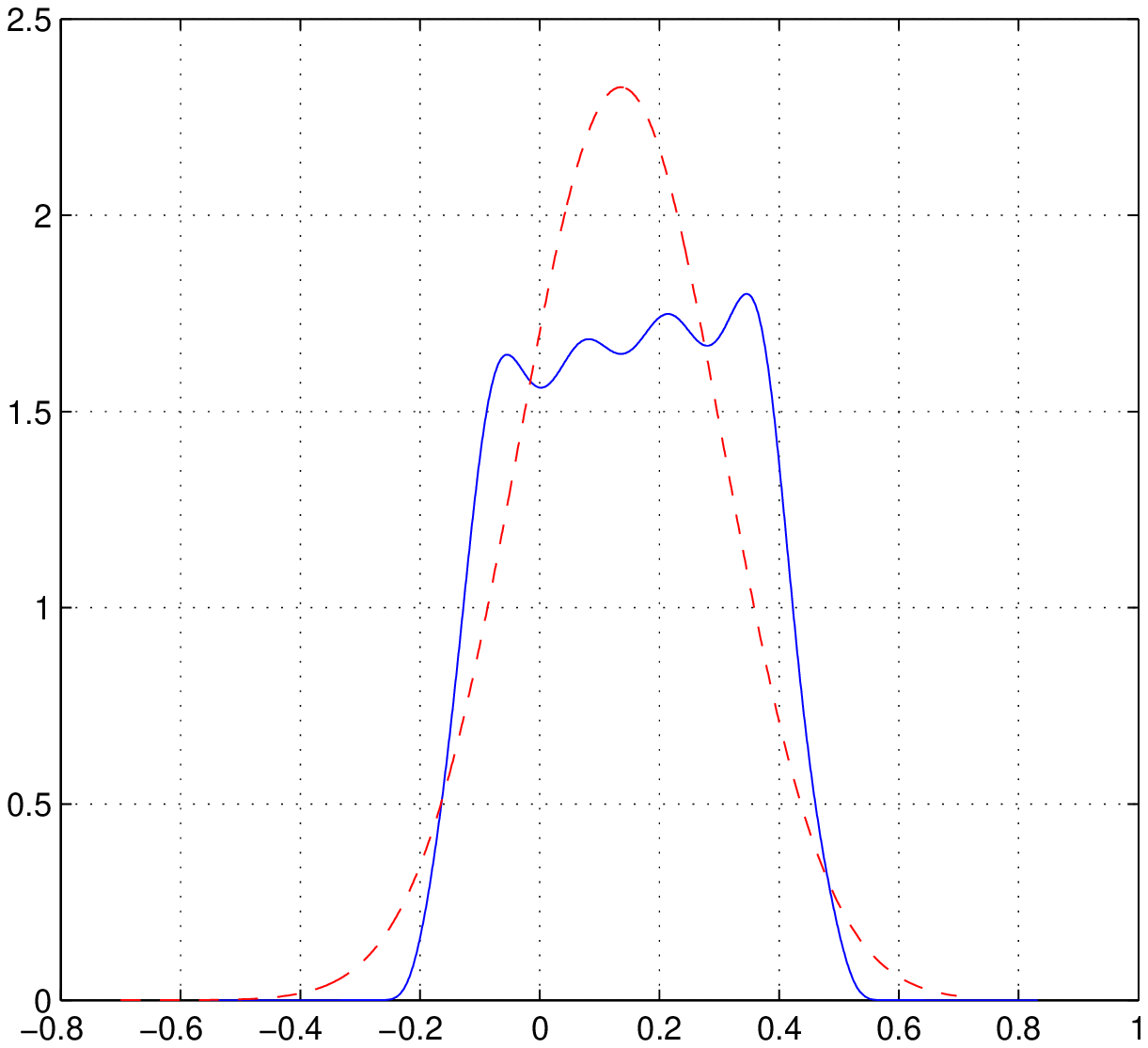}
\includegraphics[scale=0.45]{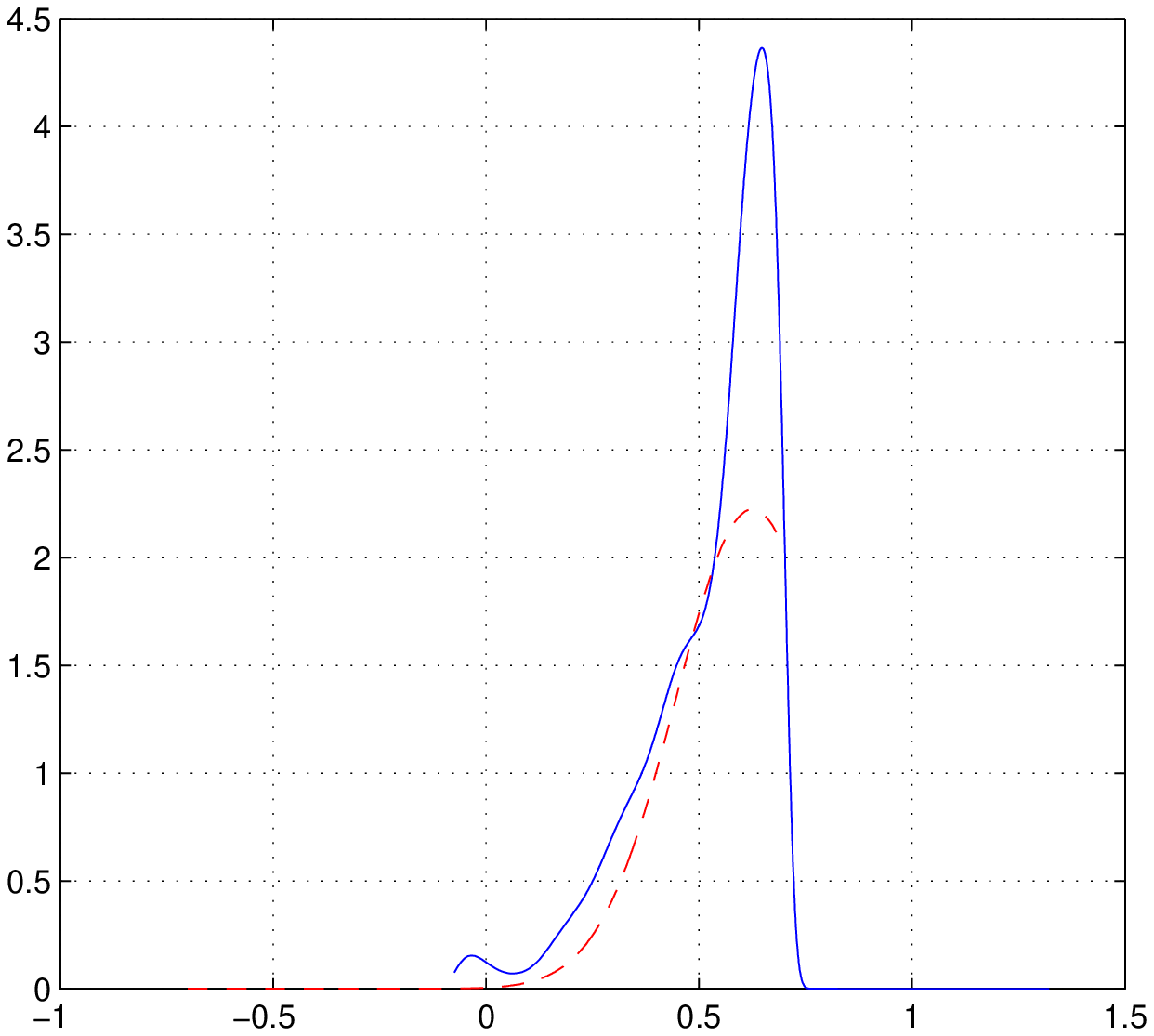}
\caption{\em  An excellent track. The first plot reports the reconstructed tracks: continuous blue line is the MIN-LOG, the dash dotted magenta line is the least squares method, and  dashed red line the effective variances. The asterisks are the track hits.
The other plots are  the PDF $P_{x_{g2}}(x,E_t,\varepsilon)$ for each hit (continuous blue lines), the dashed red lines are the approximate gaussian PDF.
 }\label{fig:figure_14}
\end{center}
\end{figure}
Figure~\ref{fig:figure_14} illustrates this easy situation in detail for a track.  The first two hits have very
narrow and very high probability
distributions (peaks around 31 and 44), and they determine
completely the track parameters in the MIN-LOG method. The minimum
search routine finds a well defined global minimum, and
the reconstructed track has parameters very near to $\beta=0$ and $\gamma=0$.
The last point has a large error but  small peak value for
$P_{x_{g2}}(x(j),E_t(j),\varepsilon-\varepsilon(j))$ (around 4.4) and a small
effective variance $\sigma_{eff}$,
it looks completely excluded from track reconstruction.
The different scales in the horizontal axis (in $cm$) and in
the vertical axis (in $\mu m$) amplify the bending of the tracks.
The approach with the effective variances reconstructs a track
as good as that of the MIN-LOG for identical reasons.
The least squares method  strongly deviates from the exact track.

The explanations of the excellent results of the MIN-LOG
based on the effective variances are misleading in some cases.
Some hits have  small effective variance but large position
errors, they are expected to pass large errors in the
reconstructed tracks with linear equations.
On the contrary, our probability distributions are very
different from gaussian distributions and
their long tails allow the existence of hits whose
positions are impossible in a gaussian model. The results
of the MIN-LOG are surprising with these events.

\subsection{Worst hits and effective hit suppression}
\begin{figure}[h]
\begin{center}
\includegraphics[scale=0.55]{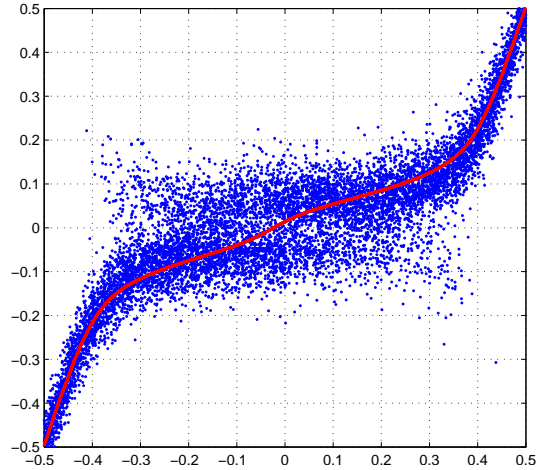}
\caption{\em Scatter plot of ( $\varepsilon$, $x_{g2}$), continuous red line $x_{g2}(\eta_2)$ (10000 points)
 }\label{fig:figure_15}
\end{center}
\end{figure}

Let us fix the limit of our approach, the natural selection
is addressed to the worst hits discussed above:
hits with a narrow effective variance and large position error.
It is easy to observe in fig.~\ref{fig:figure_15} two groups
of hits with $x_{g2}>0.1$ and $\varepsilon<-0.2$, and $x_{g2}<-0.1$
and $\varepsilon>0.2$ that belong to the sought set.
They are well known because they form the long tails of the error
distribution of the $\eta_2$ algorithm. In fact, the $\eta_2(j)$-positions are
given by the intersection of the continuous red line and straight lines
of constant $x_{g2}$-values, thus the first group has $\eta_2>0.3$
and the second group has $\eta_2<-0.3$ with position errors
greater than $0.5$ (more than 30 $\mu m$). A pure COG algorithm is
slightly better for these hits, but it is worst for
all the hits near to the red line (the large majority).
The dangerous effects of the high errors hits (outliers) are well known and
dedicated algorithms are used to attenuate their
effects. We also expected large distortions of the tracks
by the special emphasis that eq.~\ref{eq:equation_14} imposes on
the higher part of the $P_{x_{g2}}(x(j),E_t(j),\varepsilon)$.

\begin{figure}[h]
\begin{center}
\includegraphics[scale=0.43]{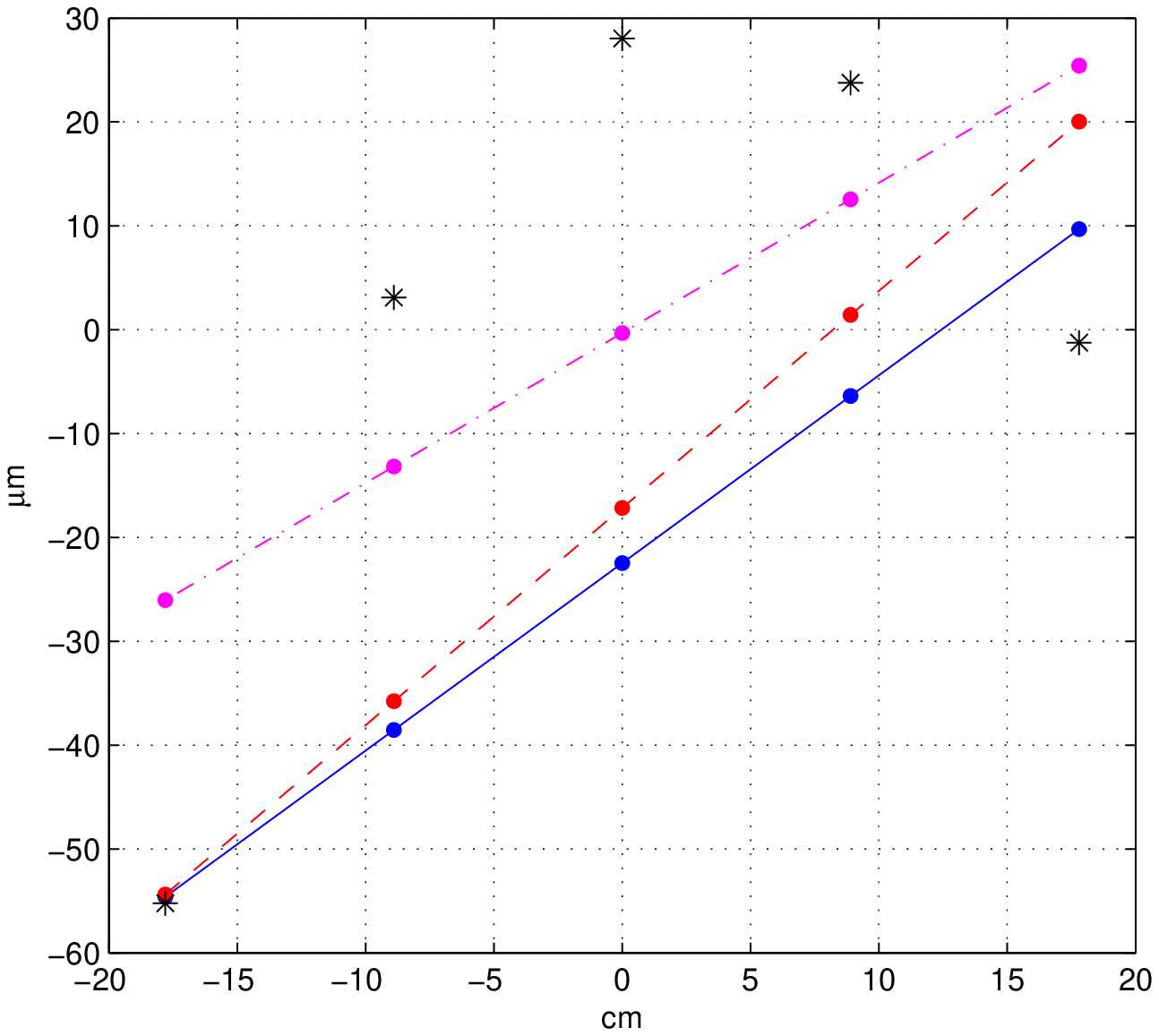}
\includegraphics[scale=0.45]{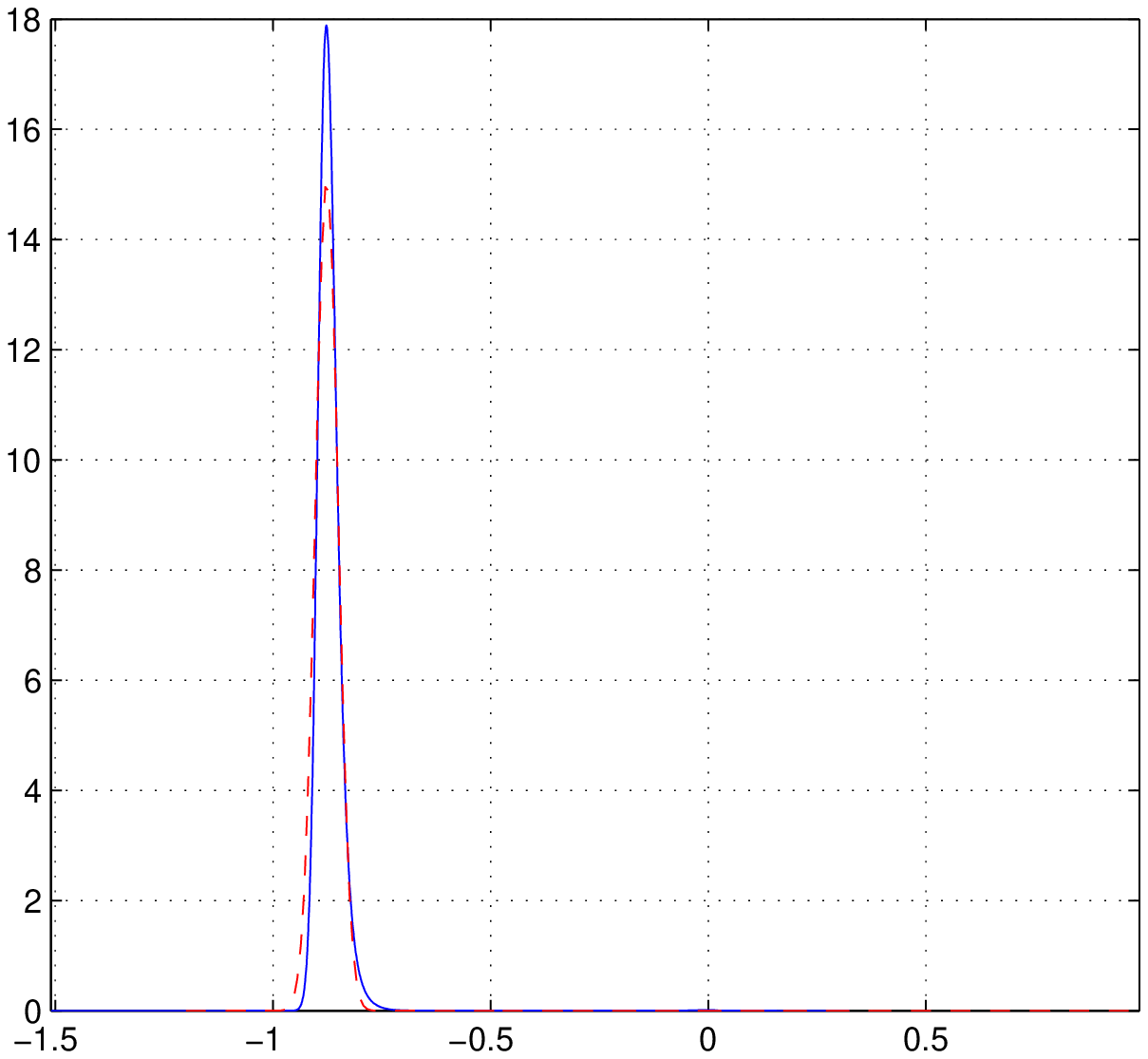}
\includegraphics[scale=0.45]{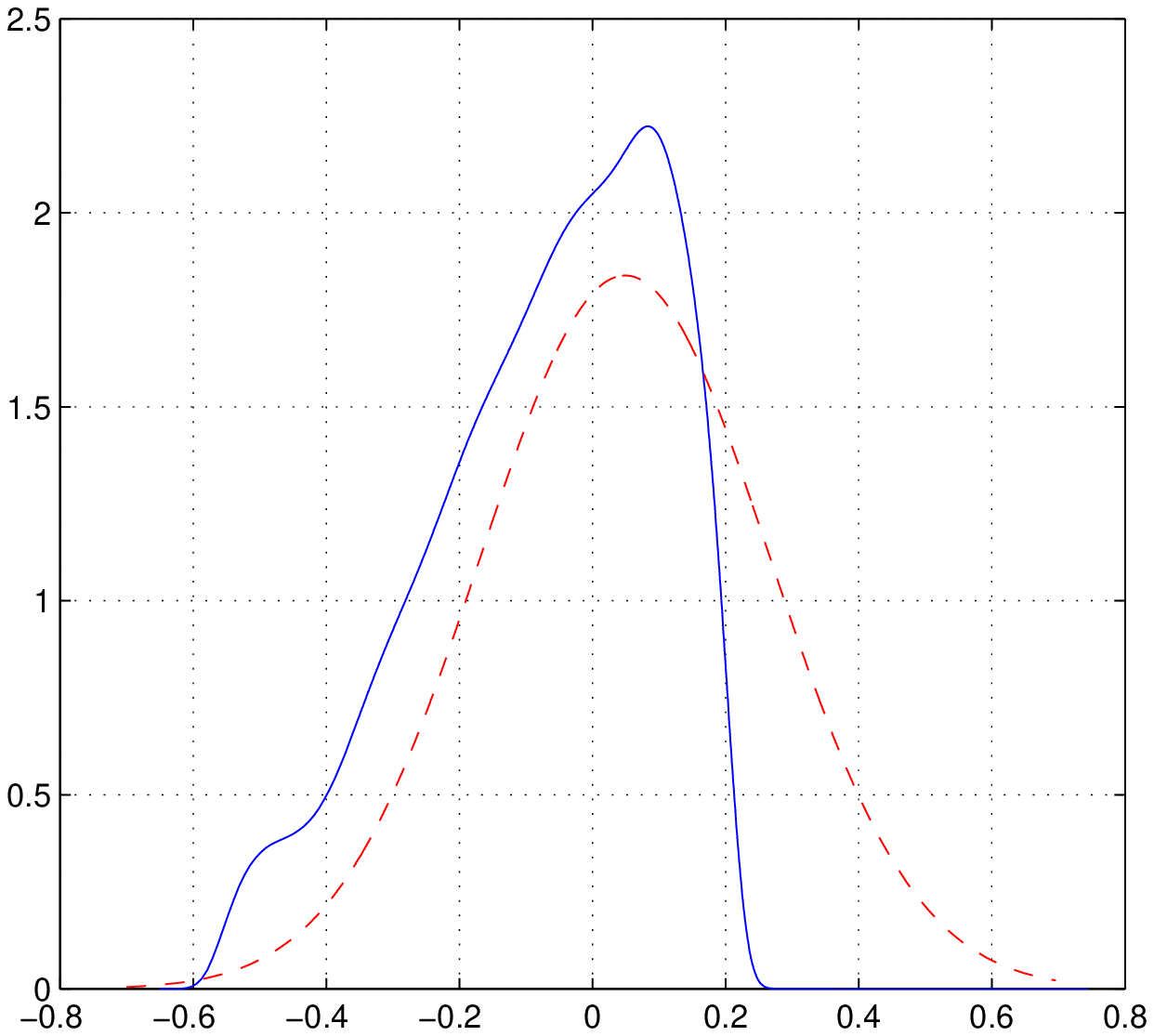}
\includegraphics[scale=0.45]{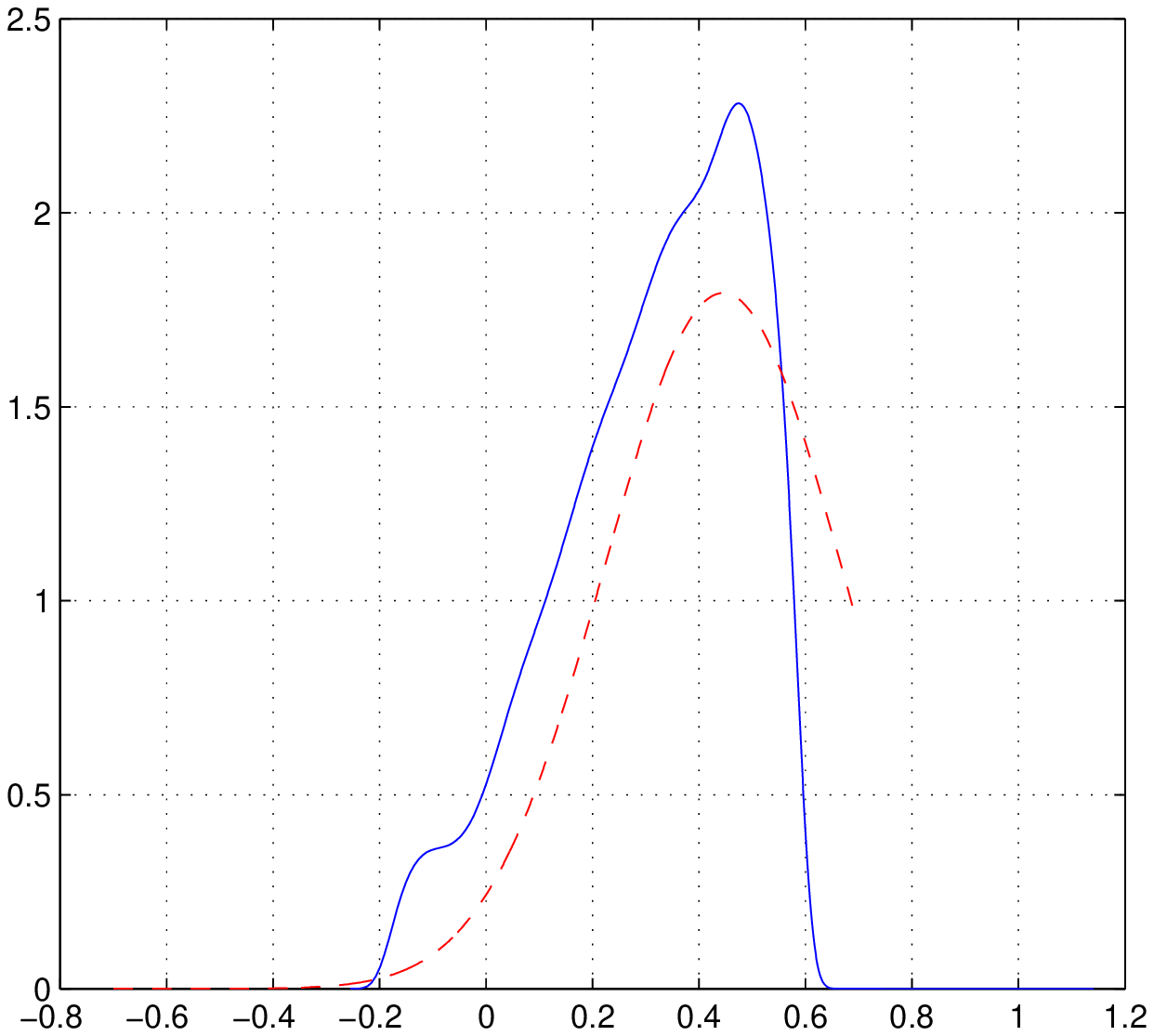}
\includegraphics[scale=0.45]{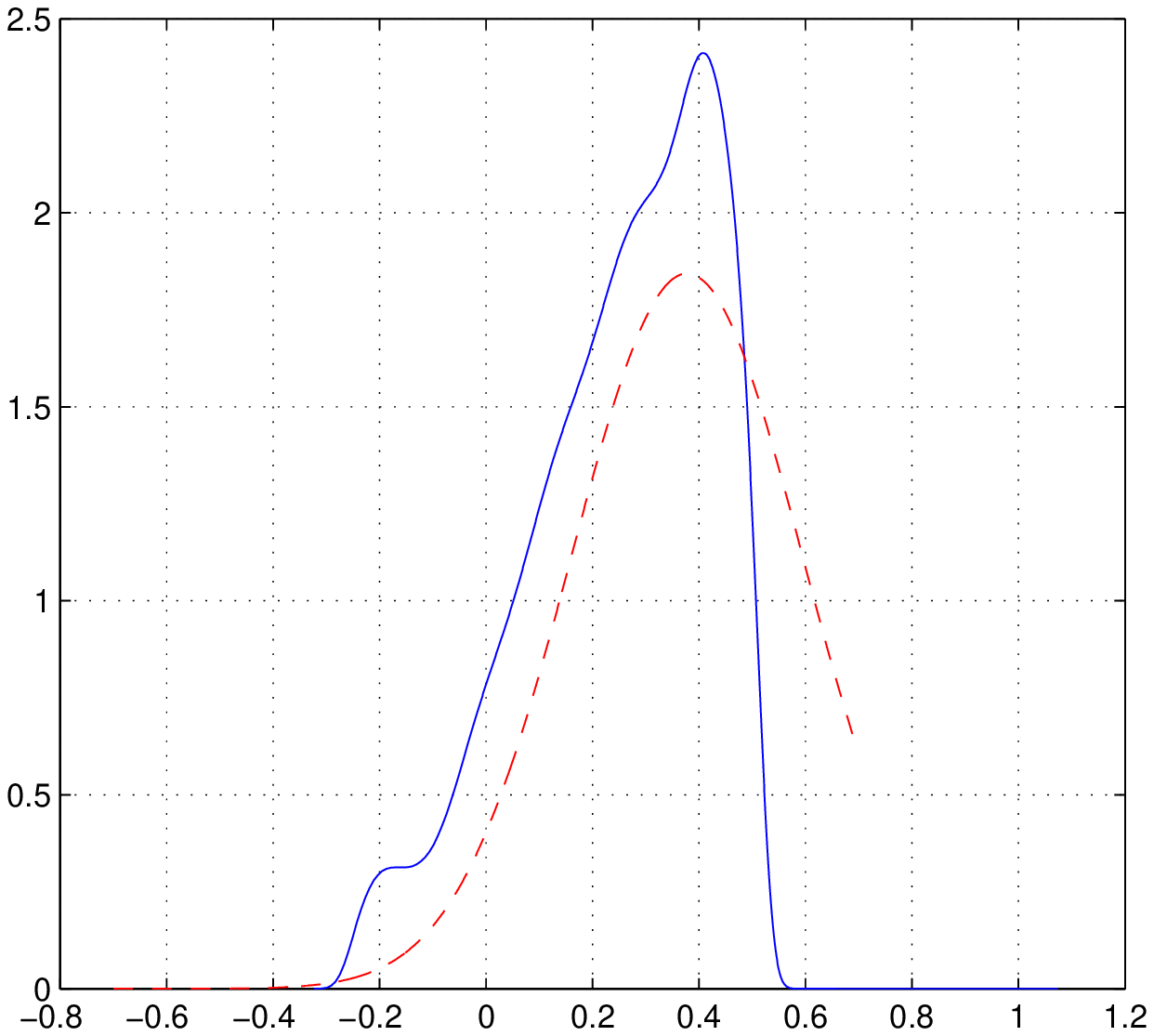}
\includegraphics[scale=0.45]{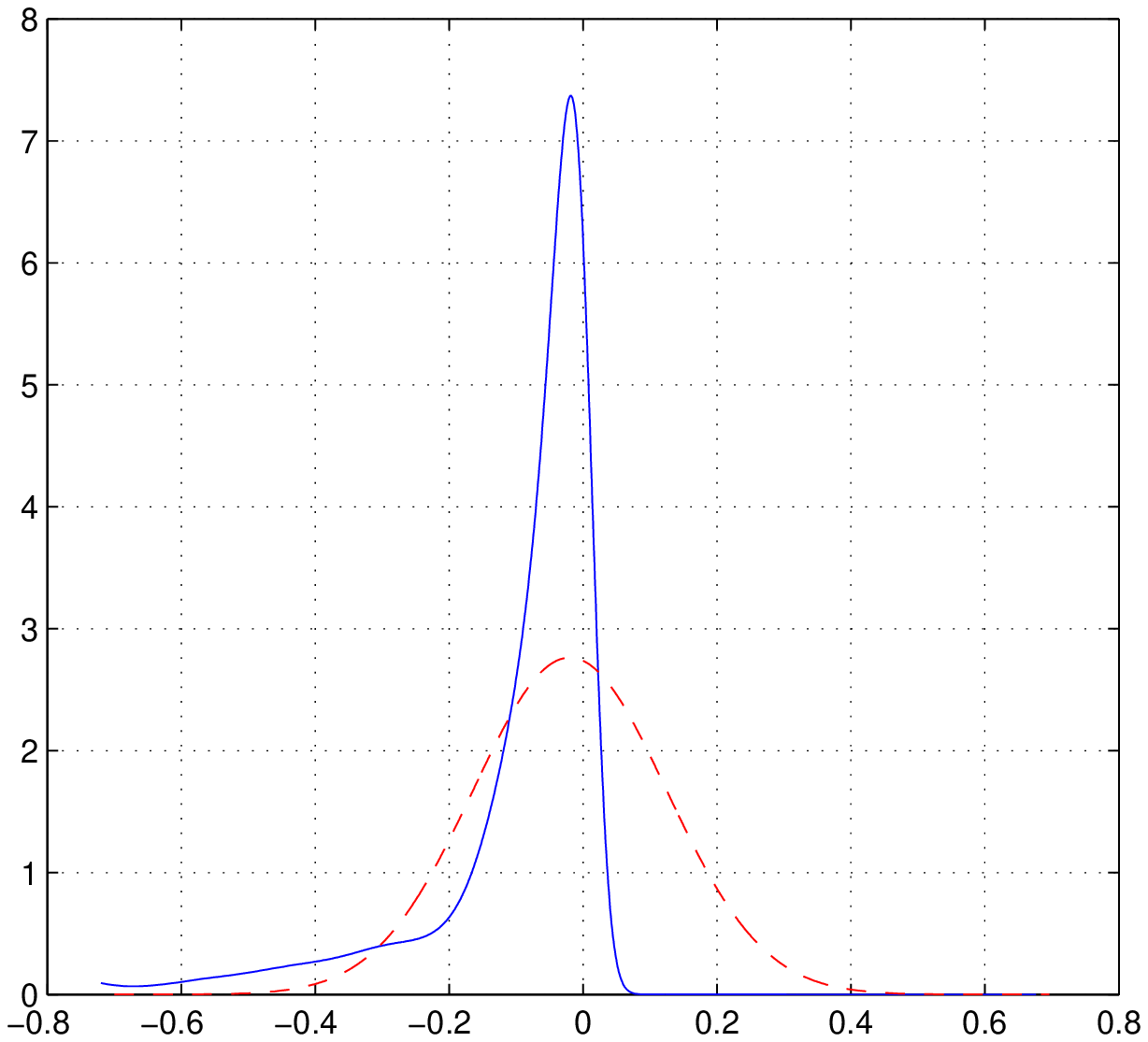}
\caption{\em  The worst track. The plot conventions are these of fig.13
 }\label{fig:figure_16}
\end{center}
\end{figure}

In fig.~\ref{fig:figure_15}, the worst hit of the worst hit-set
is that with $\varepsilon>0.4$ and $x_{g2}<-0.3$. It
has $\varepsilon=0.43$, $x_{g2}=-0.31$ and $\eta_2=-0.42$ and its
$\eta_2(j)-\varepsilon_j$. In our definition of a track, it is shifted by $-0.85$ from zero (its exact position).
The maximum of the probability
distribution is in $\eta_2(j)-\varepsilon_j=-0.85$ and the probability is almost all concentrated around
this point with a small effective variance
(fig.~\ref{fig:figure_16}).

The first run of minimum search on eq.~\ref{eq:equation_14} gives
track parameters that strongly deviate
from their exact values $(\gamma=0,\beta=0)$.
The high probability of large shifts of the first point
moves the track, the continuous blue line of
fig.~\ref{fig:figure_16}, far from its true position.
Similarly for the effective variance method, the small $\sigma_{eff}$ obliges
the track (dashed red line) to pass very near to the first point.
These two results are easy to grasp simply by observing the forms of the  PDF.
The (superficial) strong similarity of the two PDF gives no chance
to obtain a different result from the minimum search routine.
A better result is given by the least squares method
(dash dotted magenta line) for its total independence from the
form of the probability distributions.

\begin{figure}[h]
\begin{center}
\includegraphics[scale=0.5]{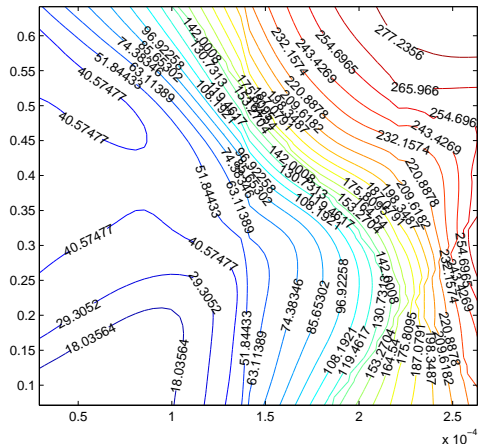}
\caption{\em The surface $L(\gamma,\beta)$ produced by the eq.4.2
around the first stop of the minimum search algorithm. The given $\beta$ and $\gamma$ are in the
center of the figure where no minimum is present.
 }\label{fig:figure_17}
\end{center}
\end{figure}

\begin{figure}[h]
\begin{center}
\includegraphics[scale=0.5]{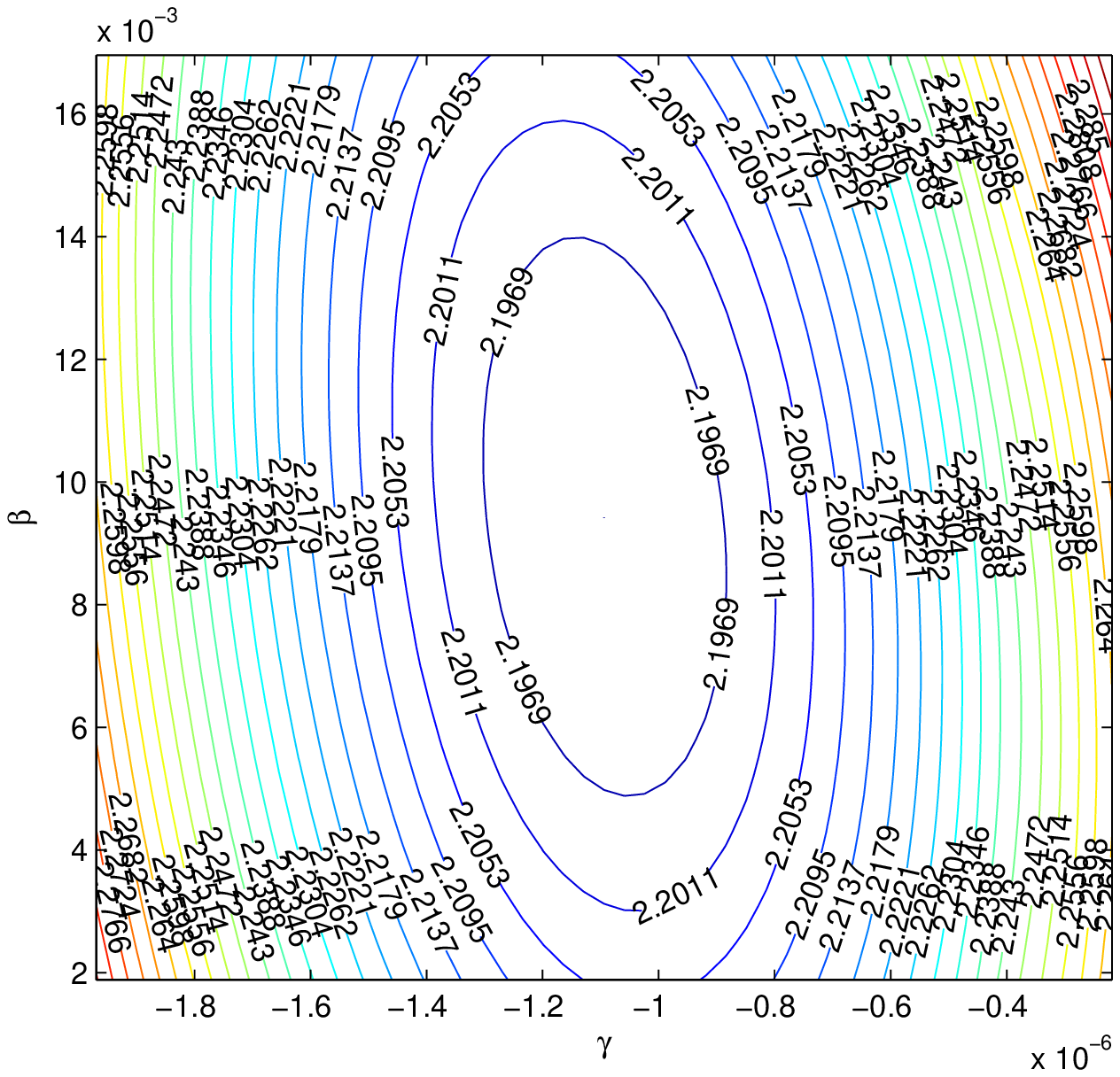}
\includegraphics[scale=0.5]{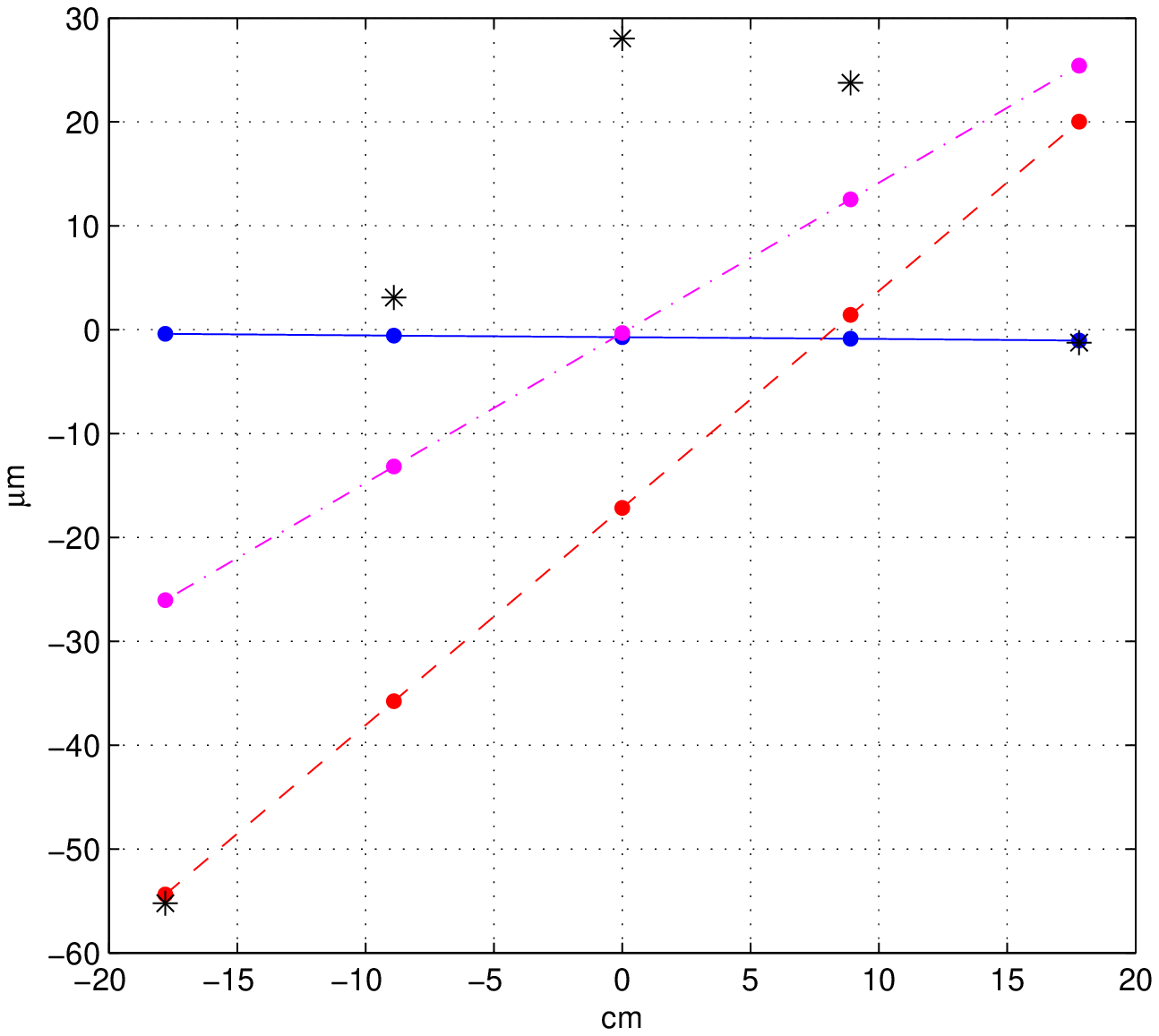}
\caption{\em To the left The surface $L(\gamma,\beta)$ produced by the eq.4.2 after the second run, the minimum is around $\{\gamma,\beta\}=\{-1.1\, 10^{-6},9\, 10^{-3}\}$. To the right the track
with the new parameters (continuous blue line) and the other unchanged.
 }\label{fig:figure_18}
\end{center}
\end{figure}

In our first random scanning of simulated tracks, we obtained good or excellent reconstructions
similar to fig.~\ref{fig:figure_14},
and few unpleasant results similar to fig.~\ref{fig:figure_16}, but not so bad. However, having in
each case a consistency with the effective variance approximation,
these bad results were considered an unavoidable defect of the method.
The consistency removed the suspicion of any connection with the well known
limitations of the minimum search routines in complex surfaces.
In fact, the minimum search routine is initialized
with the parameters obtained with the effective variances and it stops after a
fixed step number.
When the map similar  (fig.~\ref{fig:figure_17})
of eq.~\ref{eq:equation_14} does not show any minimum for these
parameters, it was natural to rerun the algorithm from this position
to look for a minimum within few other steps (this is an expected
imprecision of the minimum search routine), but still consistent
with the effective variance result.
On the contrary, the additional run gives the outputs of
fig.~\ref{fig:figure_18}. Here the track parameters are
almost exactly around zero as our ideal configuration, and a nice minimum is evident.
This result is very surprising because large
error hits induces always strong deviation to a track.
The first result looked absolutely reasonable, and
complications from these high noise hits are expected and very
difficult to isolate. Dedicated algorithms some times are able to attenuate
these effects (outliers suppression), some times no.
In the absence of these dedicated tools, we expected negligible
corrections from the second
run of the minimum search. But, without any special algorithm,
our worst track becomes the best of the three illustrated.

To exclude the possibility of a lucky accident, tracks containing other supposed
worst hits were explored. Similar
results were obtained at the small price of additional runs of the minimum search routine.
We tested even the stability to a slight perturbation of
the probability distributions, imprecisions or drift of the $a_j$-functions
must always be considered.
The minima of eq.~\ref{eq:equation_14} for tracks containing these worst hits move slightly,
but they remain well positioned around their excellent values.

\subsection{Tentative explanation}

\begin{figure}[b]
\begin{center}
\includegraphics[scale=0.45]{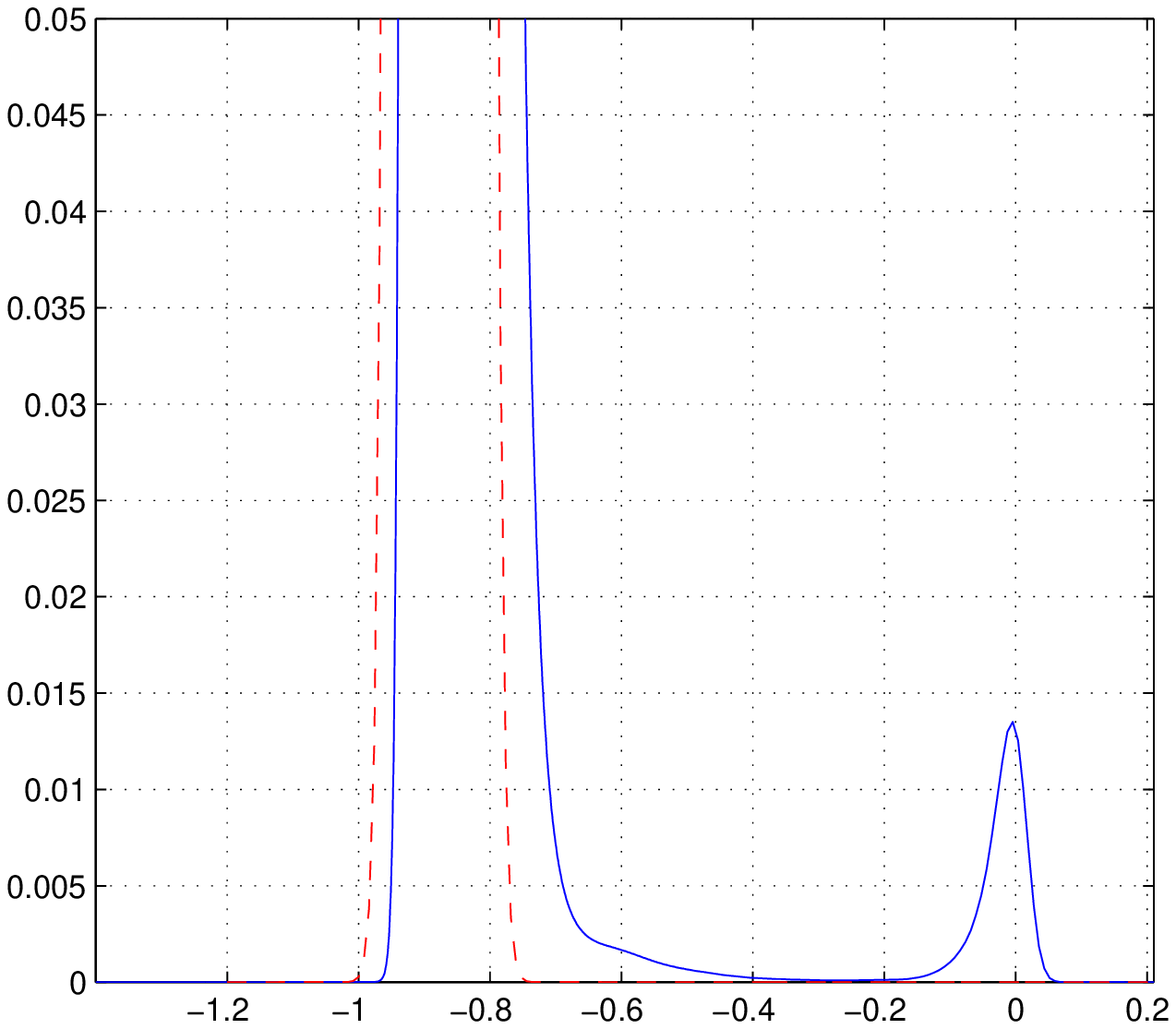}
\includegraphics[scale=0.45]{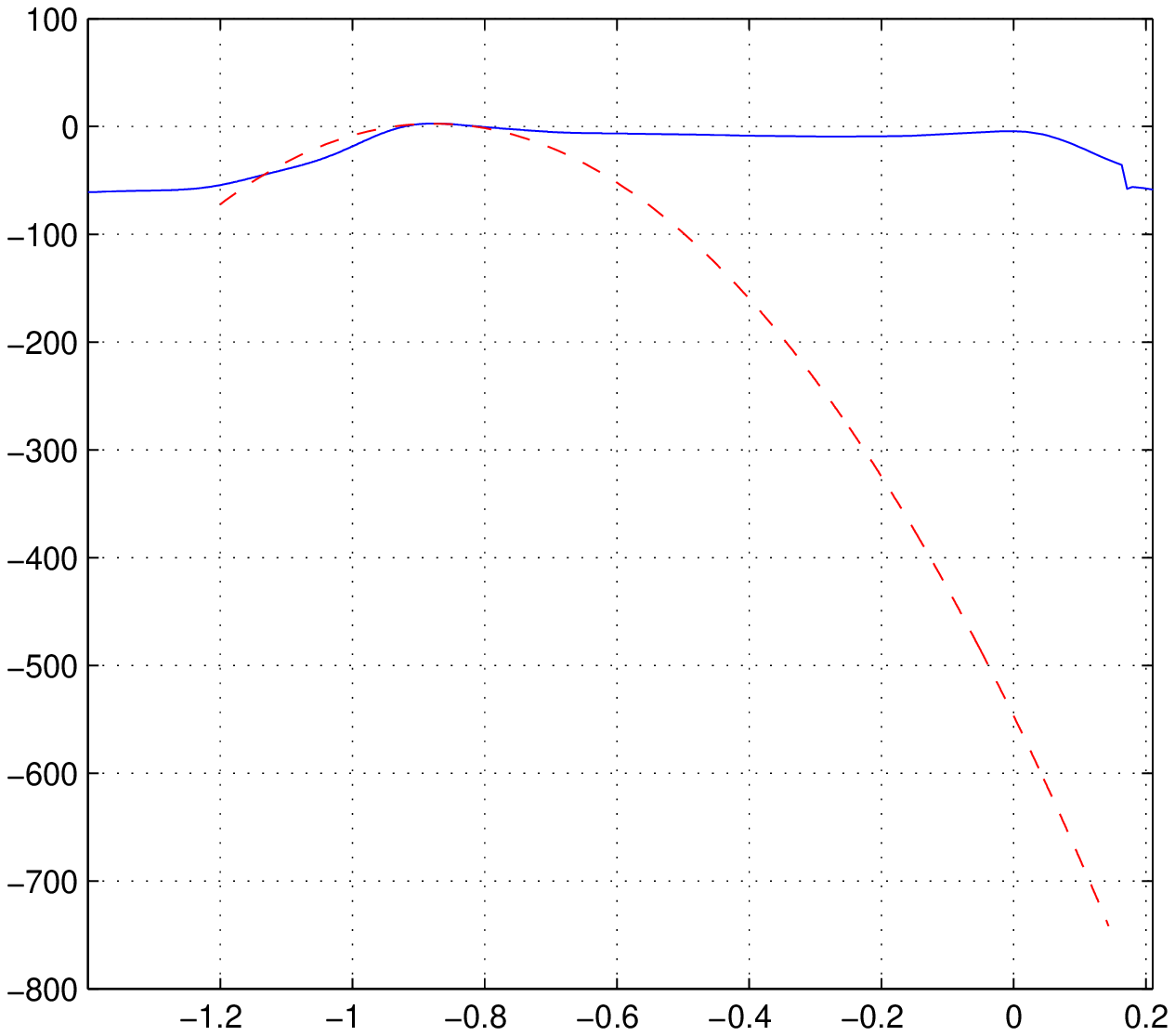}
\caption{\em To the left, the tails of the PDF for the worst hit in linear scale
(continuous blue line our PDF,
dashed red line effective gaussian). To the right the PDF
in logarithmic scale.
 }\label{fig:figure_19}
\end{center}
\end{figure}

The full explanation of these effective hit selections is not easy.
It has to do with non linear problems that
depend in an essential way from the all the other hits in the track.
The long range of our PDF  (eq.~\ref{eq:equation_13}) is implied, the tails give a
non negligible probability even to far placed points. In the
build up of the maximum for the likelihood, the function
$P_{x_{g2}}(x(j),E_t(j),\varepsilon)$ for the worst hit does not go to
zero so rapidly to exclude any maximum near
to other hits. On the contrary, a narrow gaussian
goes to zero too fast and forces the maximum to be near to its center.
Figure~\ref{fig:figure_19} clearly illustrates the
large differences of the effective gaussian and $P_{x_{g2}}(x(j),E_t(j),\varepsilon)$ for this case,
in fig.~\ref{fig:figure_16} they look similar.
As shown in the left part of fig.~\ref{fig:figure_19}, $P_{x_{g2}}(x(j),E_t(j),\varepsilon)$
has a long tail and a small peak around zero.
The presence of a second peak is a characteristic property of
eq.~\ref{eq:equation_0} as illustrated in fig.~\ref{fig:figure_2}.
At increasing $\varepsilon$, the functions $\{a_j(\varepsilon)\}$ drive
the lowest bump of fig.~\ref{fig:figure_2} to
lower values of $x_{g2}$ and decrease its height reproducing
a fading cloud of hits at negative $x_{g2}$.
The constant $x_{g2}$ plane of fig.~\ref{fig:figure_19}
cuts the lowest part of this set of bumps producing a small peak.
For this, the track parameters obtained with the insertion
of worst hit are slightly better than these obtained without.
The main result is produced  by the remaining four hits, and they simulate
the suppression of the worst hit. It would be nice if this type of
hit suppressions would be effective even with the
fake hits.  In any case, it is evident the advantage of using well tuned PDF.

After the isolation of this surprising event, we found other types of strange events,
there the minima of eq.~\ref{eq:equation_14} look to know the exact answer with the other two methods being in the dark.
Thus, our good intention to present a worst track reconstruction can not be concluded.
The set of data we expected to be a worst case turns
out to be good or excellent.
Truly bad tracks look to be produced when three or
more hits have some fake alignment, the track parameters tend to be polarized
in that direction giving an unsatisfactory result, but consistent with least squares method.

\subsection{Different Noise Amplitudes}

To test the stability of our approach and the characteristics of the worst hits,
we scale the saved random numbers  of the
additive noise and explore the track reconstruction with different noise amplitudes.
All the other parameters of the simulation are kept fixed,
thus we can follow the noise effect in each track and the modifications
of distributions of the track parameters. The noise values explored
are 4, 6, 8, 10, 12, 14 ADC counts ( or better for an average signal to noise
ratio ($S/N$) of 40, 27, 20, 16, 13, 11). The FWHM of the distributions
for $\gamma$ and $\beta$ increases with increasing the noise, as
expected, but the ratios of the FWHM for the two approaches
(MIN-LOG and least squares) are almost constant with a slight
degradation at increasing noise.

We see that the track of fig.~\ref{fig:figure_18} starts to assume its
form at a noise level of 6 ADC counts ($S/N=27$).
The effective variance method gives a track that deviates largely
from its exact position just at this noise level.
The least
squares method gives a bad reconstruction for any noise,
and our MIN-LOG is almost exact for
any noise. The large error
of the first hit is due to the two strip COG algorithm, in some
case, the noise increases the value of the left strip beyond
that of the right strip giving a change of sign to $x_{g2}$.
This type of error is rare for noise up to 8 ADC counts ($S/N=20$),
but become frequent at higher noise.

\subsection{Confronts and generalizations}

The exploration of two types of sensors, a single direction and no magnetic field, is
a small fraction of all the conditions encountered in real experiments,
but it is impossible to cover additional configurations especially in the absence of true data.
In any case, we can extract some indications for other configurations.

One evident outcome of this approach is the relation of the COG systematic
errors with the effective variances.
Figures~\ref{fig:figure_12} and~\ref{fig:figure_15} illustrate
clearly this condition, the COG systematic errors increase when
the red line of fig.~\ref{fig:figure_15} nears constant $x_{g2}$ lines
and these regions have high effective variances as shown
in fig.~\ref{fig:figure_12}. Thus, the exploration of the
systematic COG error can give some hints of the probable
improvements given the use our PDF respect
to the least squares. The trends of the COG errors were carefully analyzed in
ref.~\cite{landi01,landi02,landi03}, and those results can be helpful even now.
In any case, to a flatter COG histogram corresponds least squares results that nears to
the MIN-LOG results, but the MIN-LOG will be better for the proper
handling of the signal amplitude.

Another results can be read in fig.~\ref{fig:figure_7} and fig.~\ref{fig:figure_12}
and their "microscopic" visualization of the $\sigma_{eff}$-distributions on the strip:
the standard  resolution estimate of ref.~\cite{hartmann} ($\sigma_x\varpropto \tau/(S/N)$)
tends to be very optimistic in the two sensor sides discussed here, confirming the general suggestion of ref.~\cite{samedov}.
In fact, the large majority of $\sigma_{eff}$-values are well above the line
at 0.03 in fig.~\ref{fig:figure_7}, and similarly in fig.~\ref{fig:figure_12} for the 0.06-line.
But, being an order of magnitude estimation, it can be an useful parameter for projects and decisions.
If this constant value is extended to a fit, the least squares results is obtained.
It is evident the superiority of our  $\sigma_{eff}$-distribution, it  contains
a lot of information essential to handle these complex physical processes, and a better
description of the strip statistical properties is transferred to the fit.

\begin{figure}[t]
\begin{center}
\includegraphics[scale=0.55]{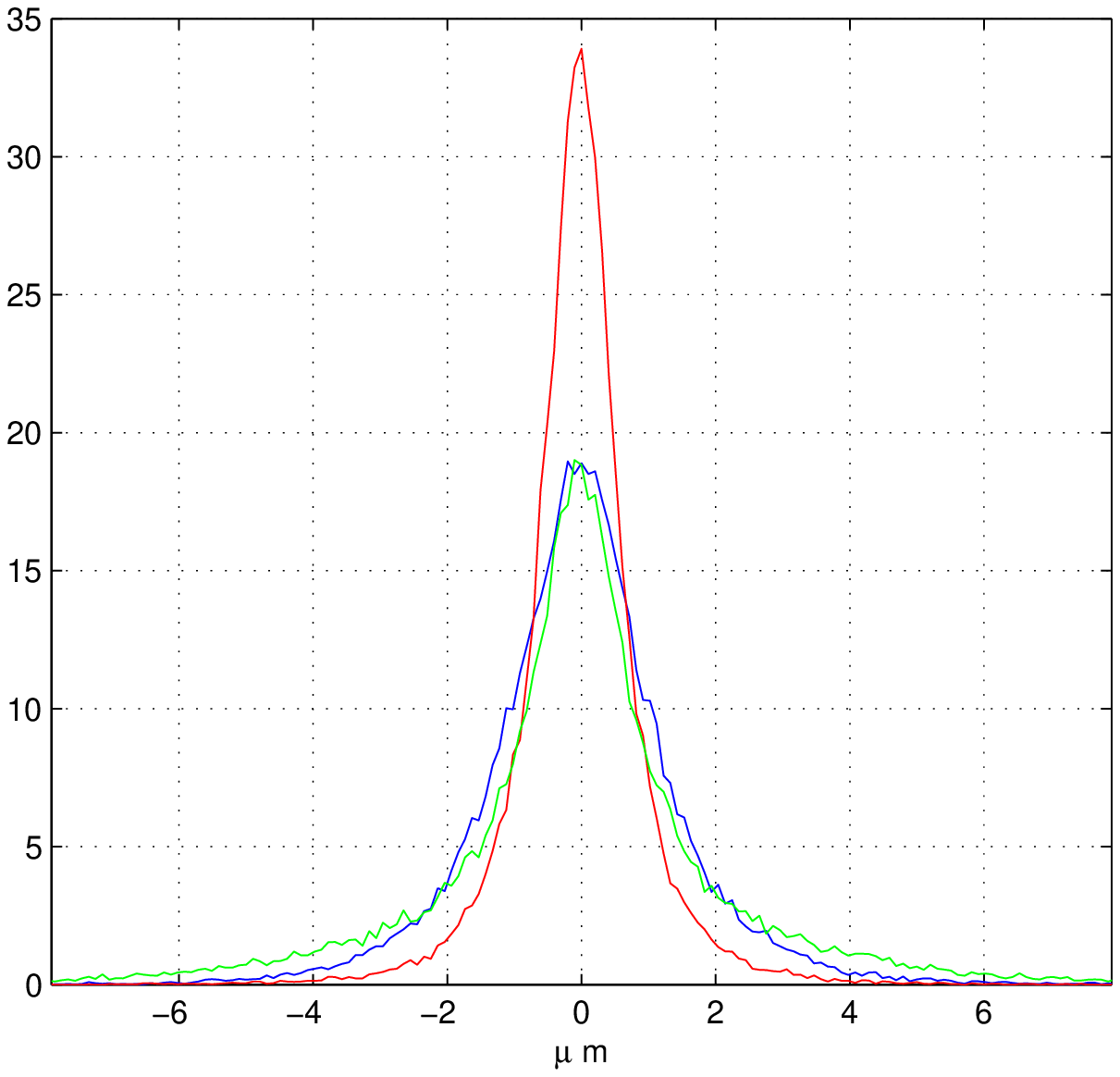}
\includegraphics[scale=0.55]{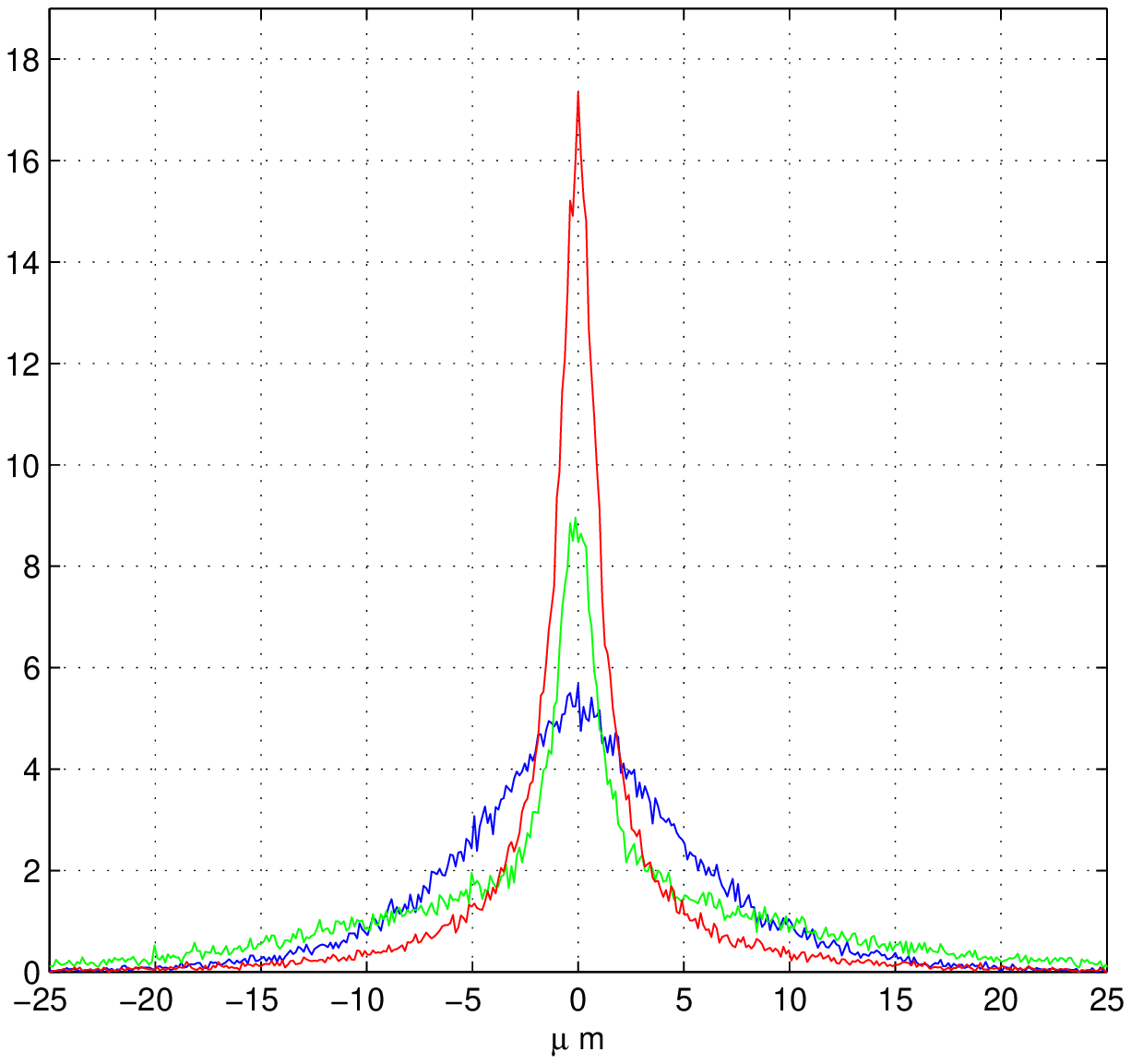}
\caption{\em Differences of the hit positions given by reconstructions and their exact values, the green line is
the distribution of the differences respect to the $\eta_2$ algorithm, the red line is for
our MIN-LOG outputs and the blue line is for the least squares. To the left, the floating strip side. To the right, the normal strip side.
 }\label{fig:figure_20}
\end{center}
\end{figure}

The gain in resolution of our procedure can be used to mitigate the reconstruction defects of the functions $\{a_j(\varepsilon)\}$,
the hit-positions, defined by the $\eta_2,\eta_3$ algorithms, introduce small  artifacts that
can be further reduced with better position determinations. Figure~\ref{fig:figure_20} illustrates
the improvement of the hit-positions given the reconstruction of the track
with our PDF compared with the input distribution given by the $\eta_2$-algorithm.
In fig.~\ref{fig:figure_20}, we also reported (with the blue line) the distributions of the differences of
the least squares respect their exact positions. These distributions turn out to be similar or worst than the
distributions of the input points. The redundancy of the track points looks to
be ineffective (or worst) in this reconstruction method.
These results, surely obtained in many other simulations, are very similar to the application
of the least squares method to points with Cauchy distributions. The  $x_{g2}$-COG, as input
for the least squares, produces difference distributions  lower than that with the input
$\eta_2$, but appreciably better than the wide and flat distribution
of the differences $(x_{g2}(j)-\varepsilon(j))$.

The production of $\delta$-rays is neglected, but we have to consider two main cases:
forward and non forward $\delta$-rays.
At orthogonal incidence, forward $\delta$-rays do not appreciably modify the signal
distribution except for an increase of the total signal collected, and this is
inserted properly. The non forward $\delta$-rays introduce a large error in the hit
reconstruction, but as for the worst tracks of section 5.2 it is probable an effective
discarding of this type of hits.

The multiple scattering effect is relevant for the low momentum tracks,
our finer details are suited for the high momentum tracks. For low momentum tracks,
the multiple scattering can be inserted with a convolution of its distribution
with our PDF for each hit.

The introduction of the magnetic field requires at least another
parameter in the minimum search of eq.~\ref{eq:equation_14}.
The plots of the minima are more difficult to produce with an additional difficulty to
follow the projections in three planes. In any case, improvements can be expected even
with the magnetic field.

\section{Conclusions}

Some aspects of our well tuned PDF are synthetically discussed with the results
of simulations of track reconstructions.
The complete forms of our analytical expressions for
the PDF can not be reported here,  they will be discussed in coming papers
with the mathematical details of their derivations.
A crucial argument is treated with the due completeness, it allows the extraction
of the average strip energies in function of the MIP impact point.
Without these special functions all our PDF would be useless.
To produce realistic simulations, we extracted
the average strip energies from test beam data with double-side silicon
strip sensors, identical to those of the PAMELA tracker.
Part of the simulated hits, based on these strip energies,
are used to produce virtual tracks for comparison of fitting methods.
The virtual track are composed by five random hits and
arranged to have identical direction and impact point.
Our PDF were completed with another set of strip energy functions
derived again from all the simulated hits. The corresponding non-linear approaches
show drastic improvements respect to the least squares method.
We find distributions of track parameters with an improvement (FWHM)
of a factor two, for the low noise side and   a factor three for
the high noise side.
The factor three of the higher noise side
is truly unexpected. We expected a result surely less than the
low noise case, we supposed the high noise a disturbance able
to render all the methods similar. This nice result
obliged us to analyze with the maximum care our outputs.
A set of tracks was explored individually to check
the consistency of the reconstructions, various details
were plotted to verify the correctness.
This deep analysis allows the isolation of another
curious effect:
an apparent hit selection that tends to avoid hits
too far from the average of the
remaining data. The long tails of our PDF
allow this suppression that
is impossible with the effective
variance or with the least squares. The approximations of
our PDF with  gaussians introduce
drastic suppressions of the tails of the distributions.
Good result can not be expected when the long
ranges with low probability are essential.
When the PDF tails are unimportant, the effective variances
turn out to be very useful, they are extracted by our PDF as parameters
for each hit (effective gaussian PDF).
The corresponding linear approximations produce a  slight degradation
of the distribution of the
track parameters, partly due to the tail suppression.
With these limitations in mind,
the effective variances can be easily
introduced in running experiments.
In many case, it is just the form of the effective
variance distributions that gives a partial explanation
of the improvements of the high noise side of the sensors.
Along the strip, very anisotropic effective-variance
distribution  is observed. An appreciable region
around the border shows very small effective
variances, hence, if a track has two hits in these regions,
the fit has an high probability to be excellent.

These results are evidently simulations built as near to the
data as possible. It would be nice
to have a dedicated test beam  to confirm them as
done in ref.~\cite{vannu} for our correction of ref.~\cite{landi03}.
With parallel tracks, the direction distributions obtained with different
approaches, can be compared to estimate the true improvements.

The use of our well tuned PDF has an high computational price and
a large set of detector parameters must be extracted from the data.
We prefer the real data, but it is possible that well calibrated
simulations could be viable substitutes. To give an idea of the
computational complexity, we produced
few thousand lines of $\mathrm{MATLAB}$~\cite{matlab} instructions
and not too less lines of $\mathrm{MATHEMATICA}$~\cite{mathematica} outputs.
Part of these developments are redundant or utilized for cross checks and
a selection of the essential elements can produce appreciable reductions.
The consistencies of these long developments are verified by
the near coincidence of the peaks of our PDF with the approximate gaussian PDF.
Here we limit to single incidence angle,
similar procedures must be done for a set of incidence angles
that cover the acceptance of the tracker.

\end{document}